\documentclass{article}
\usepackage{graphicx}
\usepackage{amsmath}
\usepackage{amsfonts}
\usepackage{amssymb}
\newtheorem{theorem}{Theorem}[section]
\newtheorem{lemma}[theorem]{Lemma}
\newtheorem{proposition}[theorem]{Proposition}
\newtheorem{corollary}[theorem]{Corollary}
\newtheorem{remark}{Remark}[section]
\newtheorem{example}{Example}[section]
\newtheorem{algorithm}{Algorithm}[section]
\numberwithin{equation}{section}
\numberwithin{figure}{section}
\numberwithin{table}{section}
\newenvironment{proof}[1][Proof]{\noindent\textbf{#1\ } }{\rule{0.5em}{0.5em}\medskip}

\begin{document}

\title{American Options under Proportional Transaction Costs: Pricing, Hedging and
Stopping Algorithms for Long and Short Positions}
\author{Alet Roux and Tomasz Zastawniak\\ \\{\small Department of Mathematics, University of York} \\{\small Heslington, York YO10 5DD, United Kingdom} \\{\small ar521@york.ac.uk, tz506@york.ac.uk}}
\date{}
\maketitle
\begin{abstract}
\noindent American options are studied in a general discrete market in the
presence of proportional transaction costs, modelled as bid-ask spreads.
Pricing algorithms and constructions of hedging strategies, stopping times and
martingale representations are presented for short (seller's) and long
(buyer's) positions in an American option with an arbitrary payoff. This
general approach extends the special cases considered in the literature
concerned primarily with computing the prices of American puts under
transaction costs by relaxing any restrictions on the form of the payoff, the
magnitude of the transaction costs or the discrete market model itself. The
largely unexplored case of pricing, hedging and stopping for the American
option buyer under transaction costs is also covered. The pricing algorithms
are computationally efficient, growing only polynomially with the number of
time steps in a recombinant tree model. The stopping times realising the ask
(seller's) and bid (buyer's) option prices can differ from one another. The
former is generally a so-called mixed (randomised) stopping time, whereas the
latter is always a pure (ordinary) stopping time.
\end{abstract}

\section{Introduction}

In this paper we study the seller's and buyer's positions in American options
when trading in the underlying asset is subject to proportional transaction
costs. The results apply to options with arbitrary payoffs in any discrete
market model and proportional transaction costs of any magnitude. We are
concerned with computing the seller's price of an American option, also known
as the upper hedging price or the ask price, as well as the buyer's price,
often referred to as the lower hedging price or the bid price. Apart from
pricing, we construct optimal strategies superhedging the positions of the
option seller and buyer, together with the respective stopping times realising
the option prices, generally a mixed (randomised) stopping time for the seller
and a pure (ordinary) stopping time for the buyer. We also consider martingale
representations for the ask and bid option prices.

The first to examine American options under proportional transaction costs in
a similar setting and level of generality as in the present paper were
Chalasani and Jha\ \cite{ChaJha01}. They established martingale
representations for options with cash settlement, subject to the simplifying
assumption that transaction costs apply at any time, except at any particular
stopping time chosen by the buyer to exercise the option. An important feature
that emerged in Chalasani and Jha's representation for the option seller's
price was the role played by mixed stopping times in place of pure stopping
times. Chalasani and Jha pointed out the non-trivial nature of computing the
option prices in their representations and the need to develop algorithms to
evaluate these prices. Our pricing algorithms solve this problem. Moreover, we
put forward algorithms for constructing the corresponding hedging strategies,
stopping times, and approximate martingales.

Bouchard and Temam \cite{BouTem05} established a dual representation for the
set of initial endowments allowing to superhedge the seller's position in an
American option in a discrete time market model with proportional transaction
costs in the setting of Kabanov, R\'{a}sonyi and Stricker \cite{KabRasStri03},
and Schachermayer \cite{Sch04}. In particular, they reproduced Chalasani and
Jha's \cite{ChaJha01} martingale representation of the seller's price.
However, note that Bouchard and Temam~\cite{BouTem05} follow a different
convention than Chalasani and Jha~\cite{ChaJha01} in that they rebalance the
portfolios in a hedging strategy before rather than after it becomes known
whether or not the American option is to be exercised.

Papers concerned with various special cases involving the hedging prices of
American options under proportional transaction costs include Koci\'{n}ski
\cite{Koc99}, \cite{Koc01}, who studied sufficient conditions for the
existence of perfectly replicating strategies for American options, Perrakis
and Lefoll \cite{PerrLef00}, \cite{PerLef04}, who investigated American calls
and puts in the binomial model, and Tokarz and Zastawniak \cite{TokZast06},
who worked with general American payoffs in the binomial model under small
proportional transaction costs.

Another group of papers, using preference-based or risk minimisation
approaches rather than superhedging for American options under proportional
transaction costs, includes Davis and Zariphopoulou \cite{DavZha95}, Mercurio
and Vorst \cite{MerVor97}, Constantinides and Zariphopoulou \cite{ConsZha01},
and Constantinides and Perrakis \cite{ConsPer04}. The work by Levental and
Skorohod \cite{LevSko97}, and Jakubenas, Levental and Ryznar
\cite{JakuLeveRyz03} shows that superhedging in continuous time leads to
unrealistic results for American options under proportional transaction costs,
thus providing motivation for exploring discrete time approaches.

The present paper complements and extends the results obtained by Chalasani
and Jha \cite{ChaJha01} and Bouchard and Temam \cite{BouTem05} by providing
pricing, hedging, stopping and approximate martingale algorithms for arbitrary
American options under proportional transaction costs in a general discrete
setting. It also extends the work on hedging prices by several of the authors
listed above, removing any restrictions imposed in the various special cases
that have been considered in the literature. As a by-product, we establish the
same martingale representations for American option prices under transaction
costs as in \cite{ChaJha01} or \cite{BouTem05} by a very different method
based on an explicit construction of the stopping times and approximate
martingales representing the ask and bid option prices. The construction
provides a geometric insight into the origin of mixed stopping times in the
seller's case. Some of the results presented here have first been established
in~\cite{Rou06}.

In the well-known case without transaction costs a stopping time that is best
for the option holder (the buyer) also happens to be the worst one for the
option writer (the seller). Similarly, a strategy hedging a shorted option is
essentially the opposite to a strategy hedging a long position in the option.
This kind of symmetry between the option seller and buyer breaks down in the
presence of transaction costs. Hedging against a stopping time that is optimal
for the buyer will generally no longer protect the seller against all other
possible exercise times. To hedge against all pure stopping times, the seller
must in effect be protected against a certain mixed stopping time. Moreover,
under transaction costs a simple relationship generally no longer exists
between strategies hedging long and short positions in the option. These
points are illustrated by the `clinical' example in Section~\ref{Sect:Example}.

In the presence of transaction costs hedging against all stopping times can
cost more than against the buyer's optimal stopping time. If the seller knew
with certainty that the option will be exercised at the buyer's optimal
stopping time, then it would only be necessary to hedge against this single
stopping time, making the seller's hedging strategy less expensive. However
the option would then no longer be of American type. This situation is
reminiscent of a Nash equilibrium.

A deeper mathematical reason behind the apparent lack of symmetry between
buyer and seller under transaction costs is that pricing for the seller as
defined by~(\ref{Eq:def-pi-a}) is a convex optimisation problem, whereas the
buyer's pricing problem~(\ref{Eq:def-pi-b}) is not of this kind, in general.
This is reflected in the pricing, hedging and stopping algorithms for the
option seller presented in this paper, which operate within the space of
convex functions (and thus have convex dual counterparts involving concave
functions), whereas the corresponding buyer's algorithms no longer act on
convex functions alone.

Computing the seller's and buyer's prices of an American option directly from
the definitions~(\ref{Eq:def-pi-a}) and (\ref{Eq:def-pi-b}) amounts to solving
large optimisation problems over the corresponding set of superhedging
strategies. Both these optimisation problems grow exponentially with the
number of time steps, as observed (for European options) by
Rutkowski~\cite{Rut98} and Chen, Sheu and Palmer~\cite{ChenPalSheu05}. In
Algorithm~\ref{Alg:seller-price1} (and the equivalent convex dual
Algorithm~\ref{Alg:seller-price2}) for the seller's price and in
Algorithm~\ref{Alg:buyer-price} for the buyer's price we present
computationally efficient dynamic programming type iterative procedures, which
grow only polynomially with the number of time steps in a recombinant tree
model. It is shown in Remark~\ref{Rem:Snell-env-extension} that
Algorithm~\ref{Alg:seller-price2} can be regarded as an extension of the
familiar Snell envelope construction to the case with transaction costs.

Numerical examples are provided to demonstrate the flexibility and efficiency
of the pricing algorithms in a realistic market model approximation. The
algorithms presented in this paper apply to options with arbitrary payoffs in
general discrete market models, including incomplete ones, with arbitrary
proportional transaction costs. The efficiency of the pricing algorithms (due
to their polynomial growth) makes it possible to cover a considerably larger
range of time steps and parameter values than in the latest numerical work by
Perrakis and Lefoll~\cite{PerLef04}, and to extend the numerical computations
beyond the binomial tree model as well as beyond puts or calls to include long
and short positions in option baskets (which are, of course, not equivalent to
a combination of puts and calls in the presence of transaction costs).

The contents of this paper are organised as follows. In
Section~\ref{Sect:Preliminaries} we fix the notation, specify the market model
with transaction costs, and present the necessary information on mixed
stopping times, approximate martingales and the families of\ functions to be
used throughout the paper. Section~\ref{Sect:Amer-opt-under-tr-costs} is the
main part of the paper. Following some definitions, pricing, hedging, stopping
and approximate martingale algorithms are presented here for both the seller
and the buyer of an American option in the presence of proportional
transaction costs, along with theorems proving the correctness of these
algorithms. A simple illustrative example, which can be followed by hand,
showing the algorithms in action can be found in Section~\ref{Sect:Example}.
In Section~\ref{Sect:num-results} we produce a number of more realistic
numerical examples. Finally, Section~\ref{Sect:Appendix} serves as an appendix
containing some technical results.

\section{Preliminaries\label{Sect:Preliminaries}}

\subsection{Market Model\label{Sect:market-model}}

Consider a finite probability space~$\Omega$ with the field~$\mathcal{F}%
=2^{\Omega}$ of all subsets of~$\Omega$, a probability measure~$Q$
on~$\mathcal{F}$ such that $Q\{\omega\}>0$ for each $\omega\in\Omega$, and a
filtration $\left\{  \emptyset,\Omega\right\}  =\mathcal{F}_{0}\subset
\mathcal{F}_{1}\subset\cdots\subset\mathcal{F}_{T}=\mathcal{F}$, the time
horizon~$T$ being a positive integer. For each $t=0,1,\ldots,T$ we denote
by~$\Omega_{t}$ the set of atoms of~$\mathcal{F}_{t}$, and identify any
$\mathcal{F}_{t}$-measurable random variable $X$ with a function defined
on~$\Omega_{t}$. We shall write $X^{\mu}$ to indicate the value of~$X$ at
$\mu\in\Omega_{t}$. Any probability measure~$P$ on~$\mathcal{F}$ can be
identified with the family of probability measures $P_{t}$ on~$\mathcal{F}%
_{t}$ such that $P_{t}(\mu)=P(\mu)$ for each $\mu\in\Omega_{t}$ and
$t=0,1,\ldots,T$.

The filtration can be represented as a tree, the atoms of~$\mathcal{F}_{t}$
corresponding to the nodes of the tree at time~$t$. We shall say that $\nu
\in\Omega_{t+1}$ is a successor node of $\mu\in\Omega_{t}$ if $\nu\subset\mu$,
this relationship corresponding to the branches of the tree. The set of
successor nodes of $\mu\in\Omega_{t}$ will be denoted by
\[
\operatorname*{succ}\mu=\{\nu\in\Omega_{t+1}\,|\,\nu\subset\mu\}.
\]

The market model consists of a risk-free bond and a risky stock. There are
proportional transaction costs on stock trades expressed as bid-ask spreads,
as in\ Jouini and Kallal~\cite{JouKal95}. Shares can be bought at the ask
price~$S_{t}^{\mathrm{a}}$ or sold at the bid price~$S_{t}^{\mathrm{b}}$,
where $S_{t}^{\mathrm{a}}\geq S_{t}^{\mathrm{b}}>0$ for each~$t=0,1,\ldots,T$,
the processes~$S^{\mathrm{a}}$ and~$S^{\mathrm{b}}$ being adapted to the
filtration. Without loss of generality, we can assume that all prices are
discounted, the bond price being~$1$ for each $t=0,1,\ldots,T$, so that a
position in bonds can be identified with cash holdings.

A portfolio $\left(  \gamma,\delta\right)  $ of cash (or bonds) and stock can
be liquidated at time~$t$ by selling stock for~$S_{t}^{\mathrm{b}}$ per share
to close a long position $\delta\geq0$ or buying stock for~$S_{t}^{\mathrm{a}%
}$ per share to close a short position $\delta<0$. The \emph{liquidation
value} of the portfolio will be%
\[
\vartheta_{t}(\gamma,\delta)=\gamma+S_{t}^{\mathrm{b}}\delta^{+}%
-S_{t}^{\mathrm{a}}\delta^{-}.
\]
The cost of setting up a portfolio $(\gamma,\delta)$ is%
\[
-\vartheta_{t}(-\gamma,-\delta)=\gamma-S_{t}^{\mathrm{b}}\delta^{-}%
+S_{t}^{\mathrm{a}}\delta^{+}.
\]

A~\emph{self-financing strategy} is a predictable process~$(\alpha_{t}%
,\beta_{t})$ representing positions in cash (or bonds) and stock at
$t=0,\ldots,T$ such that%
\begin{equation}
\vartheta_{t}(\alpha_{t}-\alpha_{t+1},\beta_{t}-\beta_{t+1})\geq0
\label{Eq:self-fin}%
\end{equation}
for each $t=0,\ldots,T-1$. The set of all self-financing strategies will be
denoted by~$\Phi$. An \emph{arbitrage opportunity} is a self-financing
strategy $\left(  \alpha,\beta\right)  \in\Phi$ such that%
\[
-\vartheta_{0}(-\alpha_{0},-\beta_{0})\leq0,\quad\vartheta_{T}(\alpha
_{T},\beta_{T})\geq0,\quad Q\left\{  \vartheta_{T}(\alpha_{T},\beta
_{T})>0\right\}  >0.
\]

It was established by Jouini and Kallal~\cite{JouKal95} that the lack of
arbitrage in the model with proportional transaction costs is equivalent to
the existence of a probability measure~$P$ on~$\Omega$ equivalent to~$Q$ and a
martingale~$S$ under~$P$ such that $S_{t}^{\mathrm{b}}\leq S_{t}\leq
S_{t}^{\mathrm{a}}$ for each $t=0,1,\ldots,T$. This result also follows from
Kabanov and Stricker~\cite{KabStr01}, Ortu~\cite{Ortu01}, Kabanov, R\'{a}sonyi
and Stricker~\cite{KabRasStri02}, \cite{KabRasStri03}, Tokarz~\cite{Tok04},
and Schachermayer~\cite{Sch04}.

\subsection{Mixed Stopping Times}

A stopping time $\tau$ is a random variable such that $\left\{  \tau
=t\right\}  \in\mathcal{F}_{t}$ for each $t=0,1,\ldots,T$. The set of stopping
times~$\tau$ with values in $\left\{  0,1,\ldots,T\right\}  $ will be denoted
by~$\mathcal{T}$. To distinguish them from mixed stopping times, defined
below, we shall sometimes refer to such $\tau$'s as \emph{pure stopping times}.

A \emph{mixed stopping time} (also called a \emph{randomised stopping time} as
in, for example, Chow, Robins and Siegmund~\cite{ChoRobSie71}, Baxter and
Chacon~\cite{BaxCha77}, or Chalasani and Jha \cite{ChaJha01}) is defined as a
non-negative adapted process~$\chi$ such that%
\[
\sum_{t=0}^{T}\chi_{t}=1.
\]
The set of all mixed stopping times will be denoted by~$\mathcal{X}$. We have
$\mathcal{T}\subset\mathcal{X}$ in the sense that each pure stopping
time~$\tau$ can be identified with a mixed stopping time~$\chi^{\tau}$ such
that for any $t=0,1,\ldots,T$%
\[
\chi_{t}^{\tau}=1_{\left\{  \tau=t\right\}  }.
\]

For any adapted process $Z$ and any mixed stopping time~$\chi$ the
\emph{time-}$\chi$\emph{ value of}~$Z$ is defined as%
\[
Z_{\chi}=\sum_{t=0}^{T}\chi_{t}Z_{t}.
\]
If $\tau$ is a pure stopping time, then $Z_{\chi^{\tau}}$ is the familiar
random variable%
\[
Z_{\chi^{\tau}}=\sum_{t=0}^{T}1_{\left\{  \tau=t\right\}  }Z_{t}=Z_{\tau}.
\]

For any mixed stopping time $\chi\in\mathcal{X}$ and any adapted process~$Z$
we define processes~$\chi^{\ast}$ and $Z^{\chi^{\ast}}$ such that for each
$t=0,1,\ldots,T$%
\begin{equation}
\chi_{t}^{\ast}=\sum_{s=t}^{T}\chi_{s},\quad Z_{t}^{\chi^{\ast}}=\sum
_{s=t}^{T}\chi_{s}Z_{s}. \label{Eq:chi-star-Z-chi-star0}%
\end{equation}
In addition, it will prove convenient to put%
\begin{equation}
\chi_{T+1}^{\ast}=0,\quad Z_{T+1}^{\chi^{\ast}}=0.
\label{Eq:chi-star-Z-chi-star1}%
\end{equation}

\subsection{Approximate Martingales}

As observed in Section~\ref{Sect:market-model}, a market model with
proportional transaction costs does not admit arbitrage if and only if there
exists a pair $(P,S)$ consisting of a probability measure~$P$ on~$\Omega$
equivalent to~$Q$ and a martingale~$S$ under~$P$ such that for each
$t=0,1,\ldots,T$%
\[
S_{t}^{\mathrm{b}}\leq S_{t}\leq S_{t}^{\mathrm{a}}.
\]
The family of such pairs $(P,S)$ will be denoted by~$\mathcal{P}$. If the
condition that $P$ should be equivalent to~$Q$ is relaxed, then the
corresponding family of pairs $(P,S)$ is to be denoted by~$\mathcal{\bar{P}}$.
The families~$\mathcal{P}$ and~$\mathcal{\bar{P}}$ can be used to represent
the prices of European options under proportional transaction costs, see
Jouini and Kallal~\cite{JouKal95}. To represent the prices of American options
we need certain larger families than~$\mathcal{P}$ or~$\mathcal{\bar{P}}$.

For any mixed stopping time~$\chi\in\mathcal{X}$ we denote by~$\mathcal{P}%
(\chi)$ the family of pairs $(P,S)$ consisting of a probability measure~$P$
on~$\Omega$ equivalent to~$Q$ and an adapted process~$S$ such that for each
$t=0,1,\ldots,T$%
\begin{gather}
S_{t}^{\mathrm{b}}\leq S_{t}\leq S_{t}^{\mathrm{a}}, \label{Eq:def-P-of-chi_a}%
\\
\chi_{t+1}^{\ast}S_{t}^{\mathrm{b}}\leq\mathbb{E}_{P}(S_{t+1}^{\chi^{\ast}%
}|\mathcal{F}_{t})\leq\chi_{t+1}^{\ast}S_{t}^{\mathrm{a}},
\label{Eq:def-P-of-chi_b}%
\end{gather}
where $\mathbb{E}_{P}$ is the expectation under~$P$. If the assumption
that~$P$ should be equivalent to~$Q$ is relaxed, the corresponding family of
pairs $(P,S)$ will be denoted by~$\mathcal{\bar{P}}(\chi)$. A pair $(P,S)$ of
this kind will be called an \emph{approximate martingale}. For a pure stopping
time $\tau\in\mathcal{T}$ we shall write $\mathcal{P}(\tau)$
and~$\mathcal{\bar{P}}(\tau)$ instead of $\mathcal{P}(\chi^{\tau})$
and~$\mathcal{\bar{P}}(\chi^{\tau})$. This notation and terminology resembles
that in Chalasani and Jha \cite{ChaJha01}.

Form Proposition~\ref{Prop:P-inclusions} we know that $\mathcal{P}%
\subset\mathcal{P}(\chi)$ and $\mathcal{\bar{P}}\subset\mathcal{\bar{P}}%
(\chi)$. It follows that the families $\mathcal{P}(\chi)$ and $\mathcal{\bar
{P}}(\chi)$ are non-empty for any $\chi\in\mathcal{X}$ in an arbitrage-free
market model, since $\mathcal{P}$ is non-empty and $\mathcal{P\subset\bar{P}}$.

\subsection{Families of Polyhedral Functions}

We denote by~$\Theta$ the family of functions $f:\mathbb{R\rightarrow
R\cup\{-\infty\}}$ such that $f\equiv-\infty$ or $f$ is an $\mathbb{R}$-valued
polyhedral function (i.e.\ continuous piecewise linear function with a finite
number of pieces).

For any $f,g$ in~$\Theta$ the maximum and minimum of $f$~and~$g$ also belong
to~$\Theta$. The \emph{epigraph} of a function $f\in\Theta$ is given by%
\[
\operatorname*{epi}f=\{(x,y)\in\mathbb{R}^{2}\,|\,x\geq f(y)\}.
\]
For any $a\geq b$ the function%
\[
h_{[b,a]}(y)=ay^{-}-by^{+}%
\]
belongs to~$\Theta$. Observe that the self-financing
condition~(\ref{Eq:self-fin}) can be written as%
\[
(\alpha_{t}-\alpha_{t+1},\beta_{t}-\beta_{t+1})\in\operatorname*{epi}%
h_{[S_{t}^{\mathrm{b}},S_{t}^{\mathrm{a}}]}.
\]

\begin{figure}
[h]
\begin{center}
\includegraphics[
height=1.6302in,
width=2.1923in
]%
{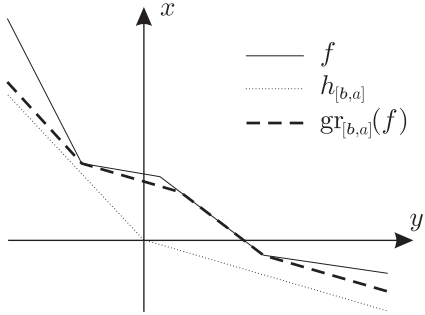}%
\caption{Gradient restriction of a function $f$ in~$\Theta$}%
\label{Fig:grad_restr4}%
\end{center}
\end{figure}

For each $f\in\Theta$ there is a unique function in~$\Theta$, denoted by
$\operatorname{gr}_{[b,a]}(f)$, such that%
\[
\operatorname*{epi}[\operatorname{gr}_{[b,a]}(f)]=\operatorname*{epi}%
h_{[b,a]}+\operatorname*{epi}f.
\]
We shall call $\operatorname{gr}_{[b,a]}(f)$ the \emph{gradient restriction}
of~$f$. This transformation is illustrated in Figure~\ref{Fig:grad_restr4}.%

If $f\in\Theta$ is a function with finite values, then it has finite limits
$f^{\prime}(+\infty)=\lim_{x\rightarrow+\infty}f^{\prime}(x)$ and $f^{\prime
}(-\infty)=\lim_{x\rightarrow-\infty}f^{\prime}(x)$. If these limits satisfy
the inequalities%
\begin{equation}
b\leq-f^{\prime}(-\infty)\quad\text{and}\quad-f^{\prime}(+\infty)\leq a,
\label{Eq:cond-for-grad-restr}%
\end{equation}
then $\operatorname{gr}_{[b,a]}(f)$ is also a function with finite values. The
financial meaning of gradient restriction is that portfolios in the epigraph
of $\operatorname{gr}_{[S_{t}^{\mathrm{b}},S_{t}^{\mathrm{a}}]}(f)$ are
precisely those that can be rebalanced in a self-financing manner at time~$t$
to yield a portfolio in the epigraph of~$f$.

Computer implementation of the three operations in~$\Theta$ mentioned above,
namely the maximum, minimum, and gradient restriction, is straightforward.
They will be used in pricing Algorithms~\ref{Alg:seller-price1}
and~\ref{Alg:buyer-price}, and in the numerical examples in
Section~\ref{Sect:num-results}.

We denote by~$\Lambda$ the family of all convex functions in~$\Theta$. It is
closed under the maximum and gradient restriction operations, but not the
minimum. For any $f\in\Lambda$ the \emph{convex dual} is defined by%
\[
f^{\ast}(x)=\inf_{y\in\mathbb{R}}(f(y)+xy)
\]
for each $x\in\mathbb{R}$. The infimum is attained whenever it is finite.
Convex duality maps~$\Lambda$ bijectively onto the family~$\Gamma$ of concave
functions $v:\mathbb{R\rightarrow R\cup\{-\infty\}}$ such that $v$ is
polyhedral (continuous piecewise linear with a finite number of pieces) on its
essential domain%
\[
\operatorname*{dom}v=\{x\in\mathbb{R}\,|\,v(x)>-\infty\}.
\]
The inverse transform from~$\Gamma$ to~$\Lambda$ is given by%
\begin{equation}
f(y)=\sup_{x\in\mathbb{R}}(f^{\ast}(x)-xy), \label{Eq:inverse-convex-duality}%
\end{equation}
with the supremum attained whenever finite.

For any $v_{1},\ldots,v_{n}\in\Gamma$ we denote by $\operatorname*{cap}%
\{v_{1},\ldots,v_{n}\}$ the \emph{concave cap} of $v_{1},\ldots,v_{n}\in
\Gamma$, defined as the smallest concave function~$v$ such that $v\geq v_{i}$
for each $i=1,\ldots,n$. It belongs to~$\Gamma$ and for each $x\in\mathbb{R}$
can be represented as%
\begin{equation}
\operatorname*{cap}\{v_{1},\ldots,v_{n}\}(x)=\max\sum_{i=1}^{n}\lambda
_{i}v_{i}(x_{i}), \label{Eq:concave-cap}%
\end{equation}
where the maximum is taken over all $\lambda_{1},\ldots,\lambda_{n}\geq0$ and
$x_{1},\ldots,x_{n}\in\mathbb{R}$ that satisfy%
\[
\sum_{i=1}^{n}\lambda_{i}=1,\quad\sum_{i=1}^{n}\lambda_{i}x_{i}=x,
\]
and $x_{i}\in\operatorname*{dom}v_{i}$ for each $i=1,\ldots,n$ such that
$\operatorname*{dom}v_{i}\neq\emptyset$, see Rockafellar~\cite{Roc97}.

Under convex duality the convex cap in~$\Gamma$ corresponds to the maximum
in~$\Lambda$,
\[
\max\{f,g\}^{\ast}=\operatorname*{cap}\{f^{\ast},g^{\ast}\}
\]
for any $f,g\in\Lambda$. The operation in~$\Gamma$ corresponding to gradient
restriction in~$\Lambda$ will be called \emph{domain restriction}. For each
$v\in\Gamma$ and each $x\in\mathbb{R}$ it is defined by%
\[
\operatorname{dr}_{[b,a]}(v)(x)=\left\{
\begin{array}
[c]{ll}%
v(x) & \text{if }x\in\lbrack b,a],\\
-\infty & \text{if }x\notin\lbrack b,a].
\end{array}
\right.
\]
For any $f\in\Lambda$ we have%
\[
\operatorname{gr}_{[b,a]}(f)^{\ast}=\operatorname{dr}_{[b,a]}(f^{\ast}).
\]
If $f\in\Lambda$ has finite values, then
\[
\lbrack-f^{\prime}(+\infty),-f^{\prime}(-\infty)]=\operatorname*{dom}f^{\ast
},
\]
and (\ref{Eq:cond-for-grad-restr}) can be written as%
\[
\lbrack b,a]\cap\operatorname*{dom}f^{\ast}\neq\emptyset.
\]
This condition guarantees that $\operatorname{gr}_{[b,a]}(f)$ has finite
values, or, equivalently, that $\operatorname{dr}_{[b,a]}(f^{\ast})$ has
non-empty essential domain.

\section{American Options under Proportional Transaction
Costs\label{Sect:Amer-opt-under-tr-costs}}

Let us take an adapted process $\left(  \xi_{t},\zeta_{t}\right)  $ with
values in $\mathbb{R}^{2}\cup\{(-\infty,-\infty)\}$ defined for all
$t=0,1,\ldots,T$ to be the payoff process of an American option. The seller of
the option must deliver to the buyer a portfolio $(\xi_{\tau},\zeta_{\tau})$
of cash and stock at an exercise time $\tau\in\mathcal{T}$ chosen by the buyer.

The pair\ $(-\infty,-\infty)$ is included among the possible values of the
payoff process to allow for the possibility that the option cannot be
exercised at certain times or nodes of the tree. This ensures that the results
of this paper are in fact valid not only for American options but also for
European or Bermudan type derivatives.

The seller can hedge a short position in the option by a self-financing
strategy $(\alpha,\beta)\in\Phi$ such that at each stopping time $\tau
\in\mathcal{T}$ he or she will be left with a solvent portfolio $(\alpha
_{\tau}-\xi_{\tau},\beta_{\tau}-\zeta_{\tau})$ once the payoff $(\xi_{\tau
},\zeta_{\tau})$ has been delivered to the buyer, that is, a portfolio such
that
\begin{equation}
\vartheta_{\tau}(\alpha_{\tau}-\xi_{\tau},\beta_{\tau}-\zeta_{\tau})\geq0.
\label{Eq:seller-superrepl-cond}%
\end{equation}
This is called a \emph{superhedging strategy for the seller}. The cost of
setting up such a strategy is $-\vartheta_{0}(-\alpha_{0},-\beta_{0})$, the
lowest of which defines the the \emph{seller's} \emph{price} (\emph{ask price,
upper hedging price}) of the option:%
\begin{equation}
\pi^{\mathrm{a}}(\xi,\zeta)=\min\left\{  -\vartheta_{0}(-\alpha_{0},-\beta
_{0})|(\alpha,\beta)\in\Phi,\forall\tau\in\mathcal{T}:\vartheta_{\tau}%
(\alpha_{\tau}-\xi_{\tau},\beta_{\tau}-\zeta_{\tau})\geq0\right\}  .
\label{Eq:def-pi-a}%
\end{equation}

On the other hand, the buyer can hedge a long position in the option by a
self-financing strategy $(\alpha,\beta)\in\Phi$ such that there is a stopping
time $\tau\in\mathcal{T}$ when he or she will be left with a solvent portfolio
$(\alpha_{\tau}+\xi_{\tau},\beta_{\tau}+\zeta_{\tau})$ after exercising the
option and receiving the payoff $(\xi_{\tau},\zeta_{\tau})$, that is, a
portfolio such that%
\begin{equation}
\vartheta_{\tau}(\alpha_{\tau}+\xi_{\tau},\beta_{\tau}+\zeta_{\tau})\geq0.
\label{Eq:buyer-superrepl-cond}%
\end{equation}
This is called a \emph{superhedging strategy for the buyer}. By setting up
such a strategy the buyer can raise the amount $\vartheta_{0}(-\alpha
_{0},-\beta_{0})$. The highest amount that can be raised in this way is called
the \emph{buyer's} \emph{price} (\emph{bid price, lower hedging price}) of the
option:%
\begin{equation}
\pi^{\mathrm{b}}(\xi,\zeta)=\max\left\{  \vartheta_{0}(-\alpha_{0},-\beta
_{0})|(\alpha,\beta)\in\Phi,\exists\tau\in\mathcal{T}:\vartheta_{\tau}%
(\alpha_{\tau}+\xi_{\tau},\beta_{\tau}+\zeta_{\tau})\geq0\right\}  .
\label{Eq:def-pi-b}%
\end{equation}

In a discrete arbitrage-free market model the minimum in~(\ref{Eq:def-pi-a})
and the maximum in~(\ref{Eq:def-pi-b}) are attained. A strategy $(\alpha
,\beta)\in\Phi$ realising the minimum in~(\ref{Eq:def-pi-a}) is referred to as
the \emph{seller's} \emph{optimal strategy}. A strategy $(\alpha,\beta)\in
\Phi$ and a stopping time $\tau\in\mathcal{T}$ realising the maximum
in~(\ref{Eq:def-pi-b}) are called the \emph{buyer's} \emph{optimal strategy}
and \emph{buyer's} \emph{optimal stopping time}.

The prices $\pi^{\mathrm{a}}(\xi,\zeta)$ and $\pi^{\mathrm{b}}(\xi,\zeta)$
provide the upper and lower bounds of the no-arbitrage interval of option
prices. Moreover, these are liquidity prices at which the option can be bought
or, respectively, sold on demand. Liquidity is important because options are
often traded as part of a strategy to hedge other derivatives.

\subsection{Seller's Case\label{Sect:sellers-case}}

\subsubsection{Seller's Pricing Algorithms}

Let $(\xi,\zeta)$ be the payoff process of an American option. For each
$t=0,1,\ldots,T$ \ and each $y\in\mathbb{R}$ we put%
\begin{equation}
u_{t}(y)=\xi_{t}+(y-\zeta_{t})^{-}S_{t}^{\mathrm{a}}-(y-\zeta_{t})^{+}%
S_{t}^{\mathrm{b}}. \label{Eq:u_t}%
\end{equation}
This defines an adapted process $u_{t}\in\Lambda$. Observe that a strategy
$(\alpha,\beta)\in\Phi$ satisfies sellers's superhedging
condition~(\ref{Eq:seller-superrepl-cond}) for a stopping time $\tau
\in\mathcal{T}$ if and only if $(\alpha_{\tau},\beta_{\tau})\in
\operatorname*{epi}u_{\tau}$.

\begin{algorithm}
\label{Alg:seller-price1}\upshape For $t=0,1,\ldots,T$ take $u_{t}\in\Lambda$
given by~(\ref{Eq:u_t}) and construct adapted processes $z_{t},v_{t},w_{t}%
\in\Lambda$ by backward induction as follows:

\begin{itemize}
\item  For each $\mu\in\Omega_{T}$ put%
\[
z_{T}^{\mu}=v_{T}^{\mu}=w_{T}^{\mu}=u_{T}^{\mu}.
\]

\item  For each $t=1,\ldots,T$ and $\mu\in\Omega_{t-1}$ put%
\begin{equation}
z_{t-1}^{\mu}=\max\{v_{t-1}^{\mu},u_{t-1}^{\mu}\}, \label{Eq:Alg_seller_max_z}%
\end{equation}
where%
\begin{align}
v_{t-1}^{\mu}  &  =\operatorname{gr}_{[S_{t-1}^{\mathrm{b}\mu},S_{t-1}%
^{\mathrm{a}\mu}]}(w_{t-1}^{\mu}),\label{Eq:Alg_seller_gr}\\
w_{t-1}^{\mu}  &  =\max\{z_{t}^{\nu}\,|\,\nu\in\operatorname*{succ}\mu\}.
\label{Eq:Alg_seller_max_w}%
\end{align}
\end{itemize}
\end{algorithm}

The resulting function $z_{0}$ will be related in
Lemma~\ref{Lem:seller-hedging} to hedging the seller's position in the
American option $(\xi,\zeta)$. In
Theorem~\ref{Thm:seller-martingale-representation} it will be shown that%
\[
\pi^{\mathrm{a}}(\xi,\zeta)=z_{0}(0).
\]

This algorithm can also be stated in terms of the dual functions%
\[
Z_{t}=z_{t}^{\ast},\quad V_{t}=v_{t}^{\ast},\quad W_{t}=w_{t}^{\ast},\quad
U_{t}=u_{t}^{\ast},
\]
which belong to~$\Gamma$. Observe that for each $t=0,1,\ldots,T$ and
$x\in\mathbb{R}$%
\begin{equation}
U_{t}(x)=\left\{
\begin{array}
[c]{ll}%
\xi_{t}+x\zeta_{t} & \text{if }x\in\lbrack S_{t}^{\mathrm{b}},S_{t}%
^{\mathrm{a}}],\\
-\infty & \text{if }x\notin\lbrack S_{t}^{\mathrm{b}},S_{t}^{\mathrm{a}}].
\end{array}
\right.  \label{Eq:dualU_t}%
\end{equation}
By the duality between $\Lambda$ and~$\Gamma$,
Algorithm~\ref{Alg:seller-price1} is equivalent to the following procedure.

\begin{algorithm}
\label{Alg:seller-price2}\upshape For $t=0,1,\ldots,T$ take $U_{t}\in\Gamma$
given by~(\ref{Eq:dualU_t}) and construct adapted processes $Z_{t},V_{t}%
,W_{t}\in\Gamma$ by backward induction as follows:

\begin{itemize}
\item  For each $\mu\in\Omega_{T}$ put%
\[
Z_{T}^{\mu}=V_{T}^{\mu}=W_{T}^{\mu}=U_{T}^{\mu}.
\]

\item  For each $t=1,\ldots,T$ and $\mu\in\Omega_{t-1}$ put%
\begin{equation}
Z_{t-1}^{\mu}=\operatorname*{cap}\{V_{t-1}^{\mu},U_{t-1}^{\mu}\},
\label{Eq:Alg_seller_max_z_dual}%
\end{equation}
where%
\begin{align}
V_{t-1}^{\mu}  &  =\operatorname{dr}_{[S_{t-1}^{\mathrm{b}\mu},S_{t-1}%
^{\mathrm{a}\mu}]}(W_{t-1}^{\mu}),\label{Eq:Alg_seller_gr_dual}\\
W_{t-1}^{\mu}  &  =\operatorname*{cap}\{Z_{t}^{\nu}\,|\,\nu\in
\operatorname*{succ}\mu\}. \label{Eq:Alg_seller_max_w_dual}%
\end{align}
\end{itemize}
\end{algorithm}

The function~$Z_{0}$ will be related in Lemma~\ref{Lem:seller-hedging} and
Remark~\ref{Rem:seller-hedging} to hedging the seller's position in the
American option $(\xi,\zeta)$. In
Theorem~\ref{Thm:seller-martingale-representation} it will be shown that%
\[
\pi^{\mathrm{a}}(\xi,\zeta)=\max_{x\in\mathbb{R}}Z_{0}(x).
\]

\begin{remark}
\upshape If the payoff is finite at some node $\mu\in\Omega_{t}$, that is,
$(\xi_{t}^{\mu},\zeta_{t}^{\mu})\in\mathbb{R}^{2}$, then at each ancestor node
$\nu\supset\mu$, where $\nu\in\Omega_{s}$ for some $s=0,1,\ldots,t$, the
functions $W_{s}^{\nu},V_{s}^{\nu},Z_{s}^{\nu}$ constructed in
Algorithm~\ref{Alg:seller-price2} have non-empty effective domains, and
$w_{s}^{\nu},v_{s}^{\nu},z_{s}^{\nu}$ in Algorithm~\ref{Alg:seller-price1}
take finite values. In particular, $z_{0}(0)$ and the maximum of $Z_{0}$ are
then finite. Indeed, for any $(P,S)\in\mathcal{P}$ it can be shown by backward
induction that the effective domains of $W_{s}^{\nu},V_{s}^{\nu},Z_{s}^{\nu}$
must contain~$S_{s}^{\nu}$. In an arbitrage-free model $\mathcal{P}$ is
non-empty, so that these effective domains must then also be non-empty.
\end{remark}

\begin{remark}
\label{Rem:Snell-env-extension}\upshape Algorithm~\ref{Alg:seller-price2} can
be viewed as a natural extension of the familiar Snell envelope construction.
In the absence of transaction costs, when $S_{t}^{\mathrm{a}}=S_{t}%
^{\mathrm{b}}=S_{t}$ for all~$t$, formula~(\ref{Eq:dualU_t}) simply defines
the cash equivalent $U_{t}=\xi_{t}+\zeta_{t}S_{t}$ of the payoff process
$(\xi_{t},\zeta_{t})$ for an American option with physical delivery,
(\ref{Eq:Alg_seller_gr_dual}) and (\ref{Eq:Alg_seller_max_w_dual}) give the
continuation value $V_{t-1}=\mathbb{E}^{\ast}(Z_{t}|\mathcal{F}_{t-1})$, where
$\mathbb{E}^{\ast}$ is the risk neutral expectation, and
(\ref{Eq:Alg_seller_max_z_dual}) becomes $Z_{t-1}=\max\{U_{t-1},V_{t-1}\}$.
\end{remark}

The workings of Algorithms~\ref{Alg:seller-price1} and~\ref{Alg:seller-price2}
will be illustrated in Example~\ref{Exl:clinical_example_new} and
Figures~\ref{Fig:seller_alg1} and~\ref{Fig:seller_alg2} in a simple two-step
binomial tree setting. The numerical results in Section~\ref{Sect:num-results}
for an American put and a bull spread in the binomial and trinomial tree
models are computed by implementing these algorithms. In a recombinant model
these computations grow only polynomially with the number of time steps,
resulting in efficient numerical work in a realistic setting.

\subsubsection{Hedging Seller's Position}

The following algorithm makes it possible to construct a strategy superhedging
a short (seller's) position in an American option with payoff process
$(\xi,\zeta)$ by starting from any portfolio in~$\operatorname*{epi}z_{0}$.

\begin{algorithm}
\label{Alg:seller-hedging}\upshape Construct a strategy $(\alpha,\beta)\in
\Phi$ by induction as follows:

\begin{itemize}
\item  Take any $\mathcal{F}_{0}$-measurable portfolio $(\alpha_{0},\beta
_{0})\in\operatorname*{epi}z_{0}$.

\item  Suppose that an $\mathcal{F}_{t}$-measurable portfolio $(\alpha
_{t},\beta_{t})\in\operatorname*{epi}z_{t}$ has already been constructed for
some $t=0,\ldots,T-1$. Since, by (\ref{Eq:Alg_seller_max_z})
and~(\ref{Eq:Alg_seller_gr}),%
\[
\operatorname*{epi}z_{t}\subset\operatorname*{epi}v_{t}=\operatorname*{epi}%
h_{[S_{t}^{\mathrm{b}},S_{t}^{\mathrm{a}}]}+\operatorname*{epi}w_{t},
\]
there is an $\mathcal{F}_{t}$-measurable portfolio $(\alpha_{t+1},\beta
_{t+1})\in\operatorname*{epi}w_{t}$ such that
\begin{equation}
(\alpha_{t}-\alpha_{t+1},\beta_{t}-\beta_{t+1})\in\operatorname*{epi}%
h_{[S_{t}^{\mathrm{b}},S_{t}^{\mathrm{a}}]}. \label{Eq:self-fin-cond-h}%
\end{equation}
Because of (\ref{Eq:Alg_seller_max_w}) we have $(\alpha_{t+1},\beta_{t+1}%
)\in\operatorname*{epi}z_{t+1}$, and since $(\alpha_{t+1},\beta_{t+1})$ is
$\mathcal{F}_{t}$-measurable, it is also $\mathcal{F}_{t+1}$-measurable,
completing the induction step.
\end{itemize}
\end{algorithm}

Because (\ref{Eq:self-fin-cond-h}) is equivalent to the self-financing
condition~(\ref{Eq:self-fin}), we know that $(\alpha,\beta)\in\Phi$. It will
be shown in Lemma~\ref{Lem:seller-hedging} that $(\alpha,\beta)$ is a
superhedging strategy for the seller.

\begin{remark}
\upshape When implementing the iterative step in
Algorithm~\ref{Alg:seller-hedging}, the portfolio $(\alpha_{t+1},\beta_{t+1})$
can be constructed from $(\alpha_{t},\beta_{t})$ as follows:

\begin{itemize}
\item  If $\alpha_{t}\geq w_{t}(\beta_{t})$, then we put $(\alpha_{t+1}%
,\beta_{t+1})=(\alpha_{t},\beta_{t})$. No rebalancing of the portfolio occurs
in this case.

\item  If $\alpha_{t}<w_{t}(\beta_{t})$, then the equation
\[
\alpha_{t}+x^{-}S_{t}^{\mathrm{b}}-x^{+}S_{t}^{\mathrm{a}}=w_{t}(\beta_{t}+x)
\]
has a solution $x$, and we put%
\[
(\alpha_{t+1},\beta_{t+1})=(\alpha_{t}+x^{-}S_{t}^{\mathrm{b}}-x^{+}%
S_{t}^{\mathrm{a}},\beta_{t}+x),
\]
which amounts to buying $x$ shares at the ask price~$S_{t}^{\mathrm{a}}$ if
$x>0$ or selling them at the bid price $S_{t}^{\mathrm{b}}$ if $x<0$. The
equation for~$x$ has a solution because $(\alpha_{t},\beta_{t})\in
\operatorname*{epi}h_{[S_{t}^{\mathrm{b}},S_{t}^{\mathrm{a}}]}%
+\operatorname*{epi}w_{t}$.
\end{itemize}
\end{remark}

The following result shows that $\operatorname*{epi}z_{0}$ can be
characterised as the set of endowments $(\gamma,\delta)$ consisting of cash
and stock that are sufficient to initiate a superhedging strategy for the seller.

\begin{lemma}
\label{Lem:seller-hedging}The following conditions are equivalent:

\begin{enumerate}
\item [$1)$]$(\gamma,\delta)\in\operatorname*{epi}z_{0}$.

\item[$2)$] There is a self-financing strategy $(\alpha,\beta)\in\Phi$ such
that $(\alpha_{0},\beta_{0})=(\gamma,\delta)$ and $(\alpha_{t},\beta_{t}%
)\in\operatorname*{epi}z_{t}$ for each $t=0,1,\ldots,T$.

\item[$3)$] There is a superhedging strategy $(\alpha,\beta)\in\Phi$ for the
seller such that $(\alpha_{0},\beta_{0})=(\gamma,\delta)$.
\end{enumerate}
\end{lemma}

\begin{proof}
$1)\Rightarrow2)$. This follows directly from the construction in
Algorithm~\ref{Alg:seller-hedging}.

$2)\Rightarrow3)$. This is so because $\operatorname*{epi}z_{\tau}%
\subset\operatorname*{epi}u_{\tau}$ by~(\ref{Eq:Alg_seller_max_z}) and the
seller's superhedging condition~(\ref{Eq:seller-superrepl-cond}) can be
written as $(\alpha_{\tau},\beta_{\tau})\in\operatorname*{epi}u_{\tau}$ for
each $\tau\in\mathcal{T}$.

$3)\Rightarrow1)$. If $(\alpha,\beta)\in\Phi$ is a strategy as in
condition~$3)$, we claim that $(\alpha_{t},\beta_{t})\in\operatorname*{epi}%
z_{t}$ for all $t=0,1,\ldots,T$. Condition~$1)$ then follows immediately. We
prove this claim by backward induction on~$t$. Since $(\alpha_{T},\beta
_{T})\in\operatorname*{epi}u_{T}=\operatorname*{epi}z_{T}$, the claim is valid
for $t=T$. Suppose that the claim holds for some $t=1,\ldots,T$, that is,
$(\alpha_{t},\beta_{t})\in\operatorname*{epi}z_{t}$. Since $(\alpha_{t}%
,\beta_{t})$ is $\mathcal{F}_{t-1}$-measurable, it follows
by~(\ref{Eq:Alg_seller_max_w}) that $(\alpha_{t},\beta_{t})\in
\operatorname*{epi}w_{t-1}$. Because the strategy is self-financing, we have
$(\alpha_{t-1}-\alpha_{t},\beta_{t-1}-\beta_{t})\in\operatorname*{epi}%
h_{[S_{t-1}^{\mathrm{b}},S_{t-1}^{\mathrm{a}}]}$. As a result, $(\alpha
_{t-1},\beta_{t-1})\in\operatorname*{epi}h_{[S_{t-1}^{\mathrm{b}}%
,S_{t-1}^{\mathrm{a}}]}+\operatorname*{epi}w_{t-1}=\operatorname*{epi}v_{t-1}$
by~(\ref{Eq:Alg_seller_gr}). Moreover, since $(\alpha,\beta)$ is a
superhedging strategy for the seller, $(\alpha_{t-1},\beta_{t-1}%
)\in\operatorname*{epi}u_{t-1}$. We can conclude
using~(\ref{Eq:Alg_seller_max_z}) that $(\alpha_{t-1},\beta_{t-1}%
)\in\operatorname*{epi}v_{t-1}\cap\operatorname*{epi}u_{t-1}%
=\operatorname*{epi}z_{t-1}$. The claim has been verified.
\end{proof}

\begin{remark}
\label{Rem:seller-hedging}\upshape By duality, since $z_{t}(y)=\sup
_{x\in\mathbb{R}}(Z_{t}(x)-xy)$, conditions~$1)$ and~$2)$ in
Lemma~\ref{Lem:seller-hedging} can be written, equivalently, as follows:

\begin{enumerate}
\item [$1^{\ast})$]$\gamma+x\delta\geq Z_{0}(x)$ for each $x\in\mathbb{R}$.

\item[$2^{\ast})$] There is a self-financing strategy $(\alpha,\beta)\in\Phi$
such that $(\alpha_{0},\beta_{0})=(\gamma,\delta)$ and $\alpha_{t}+x\beta
_{t}\geq Z_{t}(x)$ for each $x\in\mathbb{R}$ and each $t=0,1,\ldots,T$.
\end{enumerate}
\end{remark}

\subsubsection{Seller's Stopping Time and Approximate Martingale}

Our aim in this section is to construct a mixed stopping time $\hat{\chi}%
\in\mathcal{X}$ together with an approximate martingale $(\hat{P},\hat{S}%
)\in\mathcal{\bar{P}}(\hat{\chi})$ so that the ask price of an American
option\ with payoff process~$(\xi,\zeta)$ can be expressed as
\begin{equation}
\pi^{\mathrm{a}}(\xi,\zeta)=\mathbb{E}_{\hat{P}}((\xi+\hat{S}\zeta)_{\hat
{\chi}}). \label{Eq:sel-price-by-chiPS}%
\end{equation}
At the same time, we shall also construct certain auxiliary adapted processes
$\hat{\lambda},\hat{p},\hat{X},\hat{Y},\hat{Z},\hat{U},\hat{V}$.

\begin{algorithm}
\label{Alg:opt-stop}\upshape Construct a mixed stopping time $\hat{\chi}%
\in\mathcal{X}$, a probability measure~$\hat{P}$ and adapted processes
$\hat{\lambda},\hat{p},\hat{S},\hat{X},\hat{Y},\hat{Z},\hat{U},\hat{V}$ by
induction as follows:

\begin{itemize}
\item  For $t=0$ there is a $\hat{Y}_{0}\in\lbrack S_{0}^{\mathrm{b}}%
,S_{0}^{\mathrm{a}}]$ such that%
\[
Z_{0}(\hat{Y}_{0})=\max_{x\in\mathbb{R}}Z_{0}(x).
\]
By~(\ref{Eq:concave-cap}), since $Z_{0}=\operatorname*{cap}\{V_{0},U_{0}\}$,
there exist $\hat{X}_{0},\hat{S}_{0}\in\lbrack S_{0}^{\mathrm{b}}%
,S_{0}^{\mathrm{a}}]$ and $\hat{\lambda}_{0}\in\lbrack0,1]$ such that%
\begin{align*}
\hat{Y}_{0}  &  =(1-\hat{\lambda}_{0})\hat{X}_{0}+\hat{\lambda}_{0}\hat{S}%
_{0},\\
\hat{Z}_{0}  &  =(1-\hat{\lambda}_{0})\hat{V}_{0}+\hat{\lambda}_{0}\hat{U}%
_{0},
\end{align*}
where%
\[
\hat{Z}_{0}=Z_{0}(\hat{Y}_{0}),\quad\hat{V}_{0}=V_{0}(\hat{X}_{0}),\quad
\hat{U}_{0}=U_{0}(\hat{S}_{0}).
\]
Moreover, we can choose and $\hat{\lambda}_{0}=0$ if $U_{0}\equiv-\infty$. We
put%
\[
\hat{\chi}_{0}=\hat{\lambda}_{0},\quad\hat{P}_{0}=1.
\]

\item  For any $t=1,\ldots,T$ suppose that $\hat{\chi}_{s},\hat{P}_{s},\hat
{S}_{s},\hat{X}_{s},\hat{Y}_{s},\hat{Z}_{s},\hat{U}_{s},\hat{V}_{s}$ such that
$\hat{S}_{s},\hat{X}_{s},\hat{Y}_{s}\in\lbrack S_{s}^{\mathrm{b}}%
,S_{s}^{\mathrm{a}}]$ have already been constructed for $s=0,1,\ldots,t-1$.
Take any node $\mu\in\Omega_{t-1}$. By~(\ref{Eq:concave-cap}), since
$W_{t-1}^{\mu}=\operatorname*{cap}\{Z_{t}^{\nu}\,|\,\nu\in\operatorname*{succ}%
\mu\}$, it follows that%
\begin{align*}
\hat{X}_{t-1}^{\mu}  &  =\sum_{\nu\in\operatorname*{succ}\mu}\hat{p}_{t}^{\nu
}\hat{Y}_{t}^{\nu},\\
V_{t-1}^{\mu}(\hat{X}_{t-1}^{\mu})  &  =\sum_{\nu\in\operatorname*{succ}\mu
}\hat{p}_{t}^{\nu}Z_{t}^{\nu}(\hat{Y}_{t}^{\nu})
\end{align*}
for some $\hat{p}_{t}^{\nu}\geq0$ and $\hat{Y}_{t}^{\nu}\in\lbrack
S_{t}^{\mathrm{b}\nu},S_{t}^{\mathrm{a}\nu}]$, where $\nu\in
\operatorname*{succ}\mu$, such that%
\[
1=\sum_{\nu\in\operatorname*{succ}\mu}\hat{p}_{t}^{\nu}.
\]
Consider two cases:

\begin{itemize}
\item  If $t<T$, for each $\nu\in\operatorname*{succ}\mu$ use
(\ref{Eq:concave-cap}) again to deduce from $Z_{t}^{\nu}=\operatorname*{cap}%
\{V_{t}^{\nu},U_{t}^{\nu}\}$ that there exist $\hat{X}_{t}^{\nu},\hat{S}%
_{t}^{\nu}\in\lbrack S_{t}^{\mathrm{b}\nu},S_{t}^{\mathrm{a}\nu}]$ and
$\hat{\lambda}_{t}^{\nu}\in\lbrack0,1]$ such that%
\begin{align*}
\hat{Y}_{t}^{\nu}  &  =(1-\hat{\lambda}_{t}^{\nu})\hat{X}_{t}^{\nu}%
+\hat{\lambda}_{t}^{\nu}\hat{S}_{t}^{\nu},\\
\hat{Z}_{t}^{\nu}  &  =(1-\hat{\lambda}_{t}^{\nu})\hat{V}_{t}^{\nu}%
+\hat{\lambda}_{t}^{\nu}\hat{U}_{t}^{\nu},
\end{align*}
where%
\[
\hat{Z}_{t}^{\nu}=Z_{t}^{\nu}(\hat{Y}_{t}^{\nu}),\quad\hat{V}_{t}^{\nu}%
=V_{t}^{\nu}(\hat{X}_{t}^{\nu}),\quad\hat{U}_{t}^{\nu}=U_{t}^{\nu}(\hat{S}%
_{t}^{\nu}).
\]
Moreover, we can choose $\hat{\lambda}_{t}^{\nu}=0$ if $U_{t}^{\nu}%
\equiv-\infty$.

\item  If $t=T$, then for each $\nu\in\operatorname*{succ}\mu$ put%
\begin{gather*}
\hat{X}_{T}^{\nu}=\hat{S}_{T}^{\nu}=\hat{Y}_{T}^{\nu},\\
\hat{Z}_{T}^{\nu}=Z_{T}^{\nu}(\hat{Y}_{T}^{\nu}),\quad\hat{V}_{T}^{\nu}%
=V_{T}^{\nu}(\hat{X}_{T}^{\nu}),\quad\hat{U}_{T}^{\nu}=U_{T}^{\nu}(\hat{S}%
_{T}^{\nu}),\\
\hat{\lambda}_{T}^{\nu}=1.
\end{gather*}
\end{itemize}

\noindent Having considered these two cases, put%
\[
\hat{\chi}_{t}^{\nu}=\hat{\lambda}_{t}^{\nu}\left(  1-\sum_{s=0}^{t-1}%
\hat{\chi}_{s}^{\nu}\right)  ,\quad\hat{P}_{t}^{\nu}=\hat{p}_{t}^{\nu}\hat
{P}_{t-1}^{\mu},
\]
completing the induction step.
\end{itemize}
\end{algorithm}

The objects constructed in Algorithm~\ref{Alg:opt-stop} are by no means
unique, and we can choose any $\hat{\chi},\hat{P},\hat{\lambda},\hat{p}%
,\hat{S},\hat{X},\hat{Y},\hat{Z},\hat{U},\hat{V}$ satisfying the above conditions.

\begin{remark}
\upshape The $\hat{p}_{t}^{\nu}$'s play the role of conditional probabilities
from which the measure $\hat{P}$ is constructed so that for any $\nu\in
\Omega_{t}$%
\[
\hat{p}_{t}^{\nu}=\hat{P}(\nu|\mathcal{F}_{t-1})=\hat{P}_{t}(\nu
|\mathcal{F}_{t-1}).
\]
We can interpret $\hat{\lambda}_{t}^{\nu}$ as the proportion of the current
option holding and $\hat{\chi}_{t}^{\nu}$ as the proportion of an initial
option holding to be exercised at node~$\nu$ at time~$t$.
\end{remark}

Let $\hat{\chi}^{\ast}$ and $\hat{S}^{\hat{\chi}^{\ast}}$ be defined in terms
of $\hat{\chi}$ and $\hat{S}$ as in (\ref{Eq:chi-star-Z-chi-star0})
and~(\ref{Eq:chi-star-Z-chi-star1}). It follows from the construction in
Algorithm~\ref{Alg:opt-stop} that for each $t=0,1,\ldots,T$%
\begin{align}
\hat{\chi}_{t}^{\ast}\hat{Y}_{t}  &  =\hat{\chi}_{t+1}^{\ast}\hat{X}_{t}%
+\hat{\chi}_{t}\hat{S}_{t},\label{Eq:Am-PS-alg5a}\\
\hat{\chi}_{t}^{\ast}\hat{Z}_{t}  &  =\hat{\chi}_{t+1}^{\ast}\hat{V}_{t}%
+\hat{\chi}_{t}\hat{U}_{t}, \label{Eq:Am-PS-alg5b}%
\end{align}
and for each $t=1,\ldots,T$%
\begin{align}
\hat{X}_{t-1}  &  =\mathbb{E}_{\hat{P}}(\hat{Y}_{t}|\mathcal{F}_{t-1}%
),\label{Eq:Am-PS-alg6a}\\
\hat{V}_{t-1}  &  =\mathbb{E}_{\hat{P}}(\hat{Z}_{t}|\mathcal{F}_{t-1}).
\label{Eq:Am-PS-alg6b}%
\end{align}
The last two equalities, in turn, imply that for each $t=0,1,\ldots,T$%
\begin{align}
\hat{\chi}_{t+1}^{\ast}\hat{X}_{t}  &  =\mathbb{E}_{\hat{P}}(\hat{S}%
_{t+1}^{\hat{\chi}^{\ast}}|\mathcal{F}_{t}),\label{Eq:Am-PS-Lem1a}\\
\hat{\chi}_{t+1}^{\ast}\hat{V}_{t}  &  =\mathbb{E}_{\hat{P}}(\hat{U}%
_{t+1}^{\hat{\chi}^{\ast}}|\mathcal{F}_{t}). \label{Eq:Am-PS-Lem1b}%
\end{align}
We can prove (\ref{Eq:Am-PS-Lem1a}) by backward induction. For $t=T$ both
sides of (\ref{Eq:Am-PS-Lem1a}) are equal to zero. Suppose that
(\ref{Eq:Am-PS-Lem1a}) holds for some $t=1,\ldots,T$. Then
by~(\ref{Eq:Am-PS-alg5a}) and (\ref{Eq:Am-PS-alg6a})%
\begin{align*}
\hat{\chi}_{t}^{\ast}\hat{X}_{t-1}  &  =\mathbb{E}_{\hat{P}}(\hat{\chi}%
_{t}^{\ast}\hat{Y}_{t}|\mathcal{F}_{t-1})=\mathbb{E}_{\hat{P}}(\hat{\chi
}_{t+1}^{\ast}\hat{X}_{t}+\hat{\chi}_{t}\hat{S}_{t}|\mathcal{F}_{t-1})\\
&  =\mathbb{E}_{\hat{P}}(\mathbb{E}_{\hat{P}}(\hat{S}_{t+1}^{\hat{\chi}^{\ast
}}|\mathcal{F}_{t})+\hat{\chi}_{t}\hat{S}_{t}|\mathcal{F}_{t-1})=\mathbb{E}%
_{\hat{P}}(\hat{S}_{t+1}^{\hat{\chi}^{\ast}}+\hat{\chi}_{t}\hat{S}%
_{t}|\mathcal{F}_{t-1})\\
&  =\mathbb{E}_{\hat{P}}(\hat{S}_{t}^{\hat{\chi}^{\ast}}|\mathcal{F}_{t-1}),
\end{align*}
completing the proof of~(\ref{Eq:Am-PS-Lem1a}). That of (\ref{Eq:Am-PS-Lem1b})
is similar and will be omitted.

Combining (\ref{Eq:Am-PS-Lem1a}) with the fact that $\hat{S}_{t},\hat{X}%
_{t}\in\lbrack S_{t}^{\mathrm{b}},S_{t}^{\mathrm{a}}]$, we obtain%
\begin{gather*}
S_{t}^{\mathrm{b}}\leq\hat{S}_{t}\leq S_{t}^{\mathrm{a}},\\
\hat{\chi}_{t+1}^{\ast}S_{t}^{\mathrm{b}}\leq\hat{\chi}_{t+1}^{\ast}\hat
{X}_{t}=\mathbb{E}_{\hat{P}}(\hat{S}_{t+1}^{\hat{\chi}^{\ast}}|\mathcal{F}%
_{t})\leq\hat{\chi}_{t+1}^{\ast}S_{t}^{\mathrm{a}}%
\end{gather*}
for each $t=0,1,\ldots,T$, concluding that $(\hat{P},\hat{S})\in
\mathcal{\bar{P}}(\hat{\chi})$.

It will be shown in Theorem~\ref{Thm:seller-martingale-representation} that
the ask (seller's) option price can indeed be represented
by~(\ref{Eq:sel-price-by-chiPS}). For now, let us note the following result.

\begin{lemma}
\label{Lem:chi_S_P_hat}The mixed stopping time $\hat{\chi}\in\mathcal{X}$
together with the approximate martingale $(\hat{P},\hat{S})\in\mathcal{\bar
{P}}(\hat{\chi})$ constructed in Algorithm~\ref{Alg:opt-stop} satisfy%
\[
z_{0}(0)=\max_{x\in\mathbb{R}}Z_{0}(x)=\mathbb{E}_{\hat{P}}((\xi+\hat{S}%
\zeta)_{\hat{\chi}}),
\]
where $z_{0}$ and $Z_{0}$ are constructed in
Algorithms~\ref{Alg:seller-price1} and~\ref{Alg:seller-price2}.
\end{lemma}

\begin{proof}
By (\ref{Eq:Am-PS-alg5b}) and (\ref{Eq:Am-PS-Lem1b}),%
\begin{align*}
z_{0}(0)  &  =\max_{x\in\mathbb{R}}Z_{0}(x)=\hat{Z}_{0}=\chi_{0}^{\ast}\hat
{Z}_{0}=\hat{\chi}_{1}^{\ast}\hat{V}_{0}+\hat{\chi}_{0}\hat{U}_{0}%
=\mathbb{E}_{\hat{P}}(\hat{U}_{1}^{\hat{\chi}^{\ast}})+\hat{\chi}_{0}\hat
{U}_{0}\\
&  =\mathbb{E}_{\hat{P}}(\hat{U}_{1}^{\hat{\chi}^{\ast}}+\hat{\chi}_{0}\hat
{U}_{0})=\mathbb{E}_{\hat{P}}(\hat{U}_{0}^{\hat{\chi}^{\ast}})=\mathbb{E}%
_{\hat{P}}(\hat{U}_{\hat{\chi}})=\mathbb{E}_{\hat{P}}((\xi+\hat{S}\zeta
)_{\hat{\chi}})
\end{align*}
as claimed.
\end{proof}

\subsubsection{Representations of Seller's Price}

The constructions in the preceding sections lead to the following
representations of the seller's price.

\begin{theorem}
\label{Thm:seller-martingale-representation}The ask (seller's) price of an
American option with payoff process $(\xi,\zeta)$ can be represented as
follows:%
\begin{align*}
\pi^{\mathrm{a}}(\xi,\zeta)  &  =z_{0}(0)=\max_{x\in\mathbb{R}}Z_{0}%
(x)=\mathbb{E}_{\hat{P}}((\xi+\hat{S}\zeta)_{\hat{\chi}})\\
&  =\max_{\chi\in\mathcal{X}}\max_{(P,S)\in\mathcal{\bar{P}}(\chi)}%
\mathbb{E}_{P}((\xi+S\zeta)_{\chi})=\max_{\chi\in\mathcal{X}}\sup
_{(P,S)\in\mathcal{P}(\chi)}\mathbb{E}_{P}((\xi+S\zeta)_{\chi}),
\end{align*}
where $z_{0}\in\Lambda$ is constructed in Algorithm~\ref{Alg:seller-price1},
$Z_{0}\in\Gamma$ in Algorithm~\ref{Alg:seller-price2}, and $\hat{\chi}%
\in\mathcal{X}$ with $(\hat{P},\hat{S})\in\mathcal{\bar{P}}(\hat{\chi})$ in
Algorithm~\ref{Alg:opt-stop}.
\end{theorem}

\begin{proof}
From the definition~(\ref{Eq:def-pi-a}) of $\pi^{\mathrm{a}}(\xi,\zeta)$ and
Lemma~\ref{Lem:seller-hedging} we have%
\begin{align*}
\pi^{\mathrm{a}}(\xi,\zeta)  &  =\min\{-\vartheta_{0}(-\gamma,-\delta
)|(\gamma,\delta)\in\operatorname*{epi}z_{0}\}\\
&  \leq\min\{-\vartheta_{0}(-\gamma,0)|(\gamma,0)\in\operatorname*{epi}%
z_{0}\}=\min\{\gamma|(\gamma,0)\in\operatorname*{epi}z_{0}\}=z_{0}(0).
\end{align*}
It follows that%
\[
\pi^{\mathrm{a}}(\xi,\zeta)\leq z_{0}(0)=\max_{x\in\mathbb{R}}Z_{0}%
(x)=\mathbb{E}_{\hat{P}}((\xi+\hat{S}\zeta)_{\hat{\chi}})
\]
by~(\ref{Eq:inverse-convex-duality}), since $Z_{0}=z_{0}^{\ast}$, and by
Lemma~\ref{Lem:chi_S_P_hat}. On the other hand, from
Proposition~\ref{Prop:EchiPS_le_hedgingcost} and the definition of
$\pi^{\mathrm{a}}(\xi,\zeta)$ we know that for every $\chi\in\mathcal{X}$\ and
every $(P,S)\in\mathcal{\bar{P}}(\chi)$%
\[
\mathbb{E}_{P}(\left(  \xi+S\zeta\right)  _{\chi})\leq\pi^{\mathrm{a}}%
(\xi,\zeta).
\]
Because $\hat{\chi}\in\mathcal{X}$\ and $(\hat{P},\hat{S})\in\mathcal{\bar{P}%
}(\hat{\chi})$, this means that%
\[
\pi^{\mathrm{a}}(\xi,\zeta)=z_{0}(0)=\max_{x\in\mathbb{R}}Z_{0}(x)=\mathbb{E}%
_{\hat{P}}((\xi+\hat{S}\zeta)_{\hat{\chi}})=\max_{\chi\in\mathcal{X}}%
\max_{(P,S)\in\mathcal{\bar{P}}(\chi)}\mathbb{E}_{P}((\xi+S\zeta)_{\chi}).
\]
Moreover, since $\mathcal{P}(\chi)\subset\mathcal{\bar{P}}(\chi)$, by
Proposition~\ref{Prop:sup-max}%
\[
\max_{(P,S)\in\mathcal{\bar{P}}(\chi)}\mathbb{E}_{P}((\xi+S\zeta)_{\chi}%
)=\sup_{(P,S)\in\mathcal{P}(\chi)}\mathbb{E}_{P}((\xi+S\zeta)_{\chi}),
\]
for each $\chi\in\mathcal{X}$, which completes the proof.
\end{proof}

\begin{corollary}
The self-financing strategy $(\hat{\alpha},\hat{\beta})\in\Phi$ constructed in
Algorithm~\ref{Alg:seller-hedging} starting from the portfolio $(\hat{\alpha
}_{0},\hat{\beta}_{0})=(\pi^{\mathrm{a}}(\xi,\zeta),0)$ is optimal for the
option seller, that is, it realises the minimum in the definition~$($%
\ref{Eq:def-pi-a}$)$ of $\pi^{\mathrm{a}}(\xi,\zeta)$.
\end{corollary}

\begin{remark}
\upshape In general, under proportional transaction costs it can happen that%
\[
\pi^{\mathrm{a}}(\xi,\zeta)>\max_{\tau\in\mathcal{T}}\max_{(P,S)\in
\mathcal{\bar{P}}(\tau)}\mathbb{E}_{P}(\xi_{\tau}+S_{\tau}\xi_{\tau}),
\]
so there is no pure stopping time $\tau\in\mathcal{T}$ such that
$\pi^{\mathrm{a}}(\xi,\zeta)=$ $\mathbb{E}_{P}(\xi_{\tau}+S_{\tau}\xi_{\tau})$
for some $(P,S)\in\mathcal{\bar{P}}(\tau)$. This can be seen in
Example~\ref{Exl:clinical_example_new}.
\end{remark}

\subsection{Buyer's Case}

\subsubsection{Buyer's Pricing Algorithm}

Given an American option with payoff process $(\xi,\zeta)$, we define an
adapted process $u_{t}\in\Theta$ such that for each $t=0,1,\ldots,T$ and
$y\in\mathbb{R}$%
\begin{equation}
u_{t}(y)=-\xi_{t}+(y+\zeta_{t})^{-}S_{t}^{\mathrm{a}}-(y+\zeta_{t})^{+}%
S_{t}^{\mathrm{b}}. \label{Eq:j_t}%
\end{equation}
Observe that a strategy $(\alpha,\beta)\in\Phi$ satisfies the buyer's
superhedging condition~(\ref{Eq:buyer-superrepl-cond}) for a stopping time
$\tau\in\mathcal{T}$ if and only if $(\alpha_{\tau},\beta_{\tau}%
)\in\operatorname*{epi}u_{\tau}$.

\begin{algorithm}
\label{Alg:buyer-price}\upshape
For $t=0,1,\ldots,T$ take $u_{t}\in\Theta$ given by~(\ref{Eq:j_t}) and
construct adapted processes $z_{t},v_{t},w_{t}\in\Theta$ by backward induction
as follows:

\begin{itemize}
\item  For every $\mu\in\Omega_{T}$ put%
\[
z_{T}^{\mu}=v_{T}^{\mu}=w_{T}^{\mu}=u_{T}^{\mu}.
\]

\item  For every $t=1,\ldots,T$ and $\mu\in\Omega_{t-1}$ put%
\begin{equation}
z_{t-1}^{\mu}=\min\{v_{t-1}^{\mu},u_{t-1}^{\mu}\}, \label{Eq:Alg_buyer_min}%
\end{equation}
where%
\begin{align}
v_{t-1}^{\mu}  &  =\operatorname{gr}_{[S_{t-1}^{\mathrm{b}},S_{t-1}%
^{\mathrm{a}}]}(w_{t-1}^{\mu}),\label{Eq:Alg_buyer_gr}\\
w_{t-1}^{\mu}  &  =\max\{z_{t}^{\nu}\,|\,\nu\in\operatorname*{succ}\mu\}.
\label{Eq:Alg_buyer_max}%
\end{align}
\end{itemize}
\end{algorithm}

Although $u_{t},v_{t},w_{t},z_{t}$ constructed here are different processes
than those in the seller's Algorithm~\ref{Alg:seller-price1}, we use the same
symbols because they play analogous roles in the buyer's case.

In Lemma~\ref{Lem:buyer-pricing} the resulting function~$z_{0}$ will be
related to hedging the buyer's position in the American option~$(\xi,\zeta)$.
In Theorem~\ref{Thm:buyer-martingale-representation} it will be shown that%
\[
\pi^{\mathrm{b}}(\xi,\zeta)=-z_{0}(0).
\]

\begin{remark}
\upshape
In contrast to the seller's case, Algorithm~\ref{Alg:buyer-price} has no
convex dual counterpart similar to Algorithm~\ref{Alg:seller-price2}. This is
because the functions $z_{t},v_{t},w_{t}\in\Theta$ are not necessarily convex
due to the minimum featuring in~(\ref{Eq:Alg_buyer_min}).
\end{remark}

\subsubsection{Hedging Buyer's Position}

The buyer of an American option~$(\xi,\zeta)$ can select both a self-financing
strategy $(\alpha,\beta)\in\Phi$ and a stopping time $\tau\in\mathcal{T}$ when
the option will be exercised. In this section we construct a strategy
$(\alpha,\beta)\in\Phi$ and a stopping time $\tau\in\mathcal{T}$ that satisfy
the buyer's superhedging condition~(\ref{Eq:buyer-superrepl-cond}) by starting
from any portfolio in $\operatorname*{epi}z_{0}$.

\begin{algorithm}
\label{Alg:buyer-hedging1}\upshape
Construct by induction a strategy $(\alpha,\beta)\in\Phi$ and a sequence of
stopping times $\tau_{t}\in\mathcal{T}$ such that%
\[
(\alpha_{t},\beta_{t})\in\operatorname*{epi}z_{t}\setminus\operatorname*{epi}%
u_{t}\quad\text{on }\{t<\tau_{t}\}
\]
for each $t=0,1,\ldots,T$ as follows:

\begin{itemize}
\item  Take any $\mathcal{F}_{0}$-measurable portfolio $(\alpha_{0},\beta
_{0})\in\operatorname*{epi}z_{0}$ and put%
\[
\tau_{0}=\left\{
\begin{array}
[c]{ll}%
0 & \text{if }(\alpha_{0},\beta_{0})\in\operatorname*{epi}u_{0},\\
T & \text{if }(\alpha_{0},\beta_{0})\notin\operatorname*{epi}u_{0}.
\end{array}
\right.
\]

\item  Suppose that an $\mathcal{F}_{t}$-measurable portfolio $(\alpha
_{t},\beta_{t})\in\operatorname*{epi}z_{t}$ and a stopping time $\tau_{t}%
\in\mathcal{T}$ have already been constructed for some $t=0,\ldots,T-1$ so
that $(\alpha_{t},\beta_{t})\in\operatorname*{epi}z_{t}\setminus
\operatorname*{epi}u_{t}$ on~$\{t<\tau_{t}\}$. It follows by
(\ref{Eq:Alg_buyer_min}) and (\ref{Eq:Alg_buyer_gr}) that%
\[
(\alpha_{t},\beta_{t})\in\operatorname*{epi}v_{t}=\operatorname*{epi}%
h_{[S_{t}^{\mathrm{b}},S_{t}^{\mathrm{a}}]}+\operatorname*{epi}w_{t}%
\quad\text{on }\{t<\tau_{t}\}.
\]
As a result, there is an $\mathcal{F}_{t}$-measurable portfolio $(\alpha
_{t+1},\beta_{t+1})$ such that%
\[%
\begin{tabular}
[c]{cl}%
$(\alpha_{t+1},\beta_{t+1})\in\operatorname*{epi}w_{t},$ $\ (\alpha_{t}%
-\alpha_{t+1},\beta_{t}-\beta_{t+1})\in\operatorname*{epi}h_{[S_{t}%
^{\mathrm{b}},S_{t}^{\mathrm{a}}]}$ & on~$\{t<\tau_{t}\}$,\\
$(\alpha_{t+1},\beta_{t+1})=(\alpha_{t},\beta_{t})$ & on $\{t\geq\tau_{t}\}$.
\end{tabular}
\]
The self-financing condition~(\ref{Eq:self-fin}) is therefore satisfied both
on $\{t<\tau_{t}\}$ and $\{t\geq\tau_{t}\}$. By (\ref{Eq:Alg_buyer_gr}) we
have $(\alpha_{t+1},\beta_{t+1})\in\operatorname*{epi}z_{t+1}$ on
$\{t<\tau_{t}\}$. Then, put%
\[
\tau_{t+1}=\left\{
\begin{array}
[c]{ll}%
\tau_{t} & \text{if }t\geq\tau_{t},\\
t+1 & \text{if }t<\tau_{t}\text{ and }(\alpha_{t+1},\beta_{t+1})\in
\operatorname*{epi}u_{t+1},\\
T & \text{if }t<\tau_{t}\text{ and }(\alpha_{t+1},\beta_{t+1})\notin
\operatorname*{epi}u_{t+1}.
\end{array}
\right.
\]
completing the induction step.
\end{itemize}

\noindent Finally put $\tau=\tau_{T}\in\mathcal{T}$.
\end{algorithm}

The self-financing strategy $(\alpha,\beta)\in\Phi$ and stopping time $\tau
\in\mathcal{T}$ constructed in Algorithm~\ref{Alg:buyer-hedging1} are shown in
Lemma~\ref{Lem:buyer-pricing} to satisfy the superhedging
condition~(\ref{Eq:buyer-superrepl-cond}) for the buyer of the American
option~$(\xi,\zeta)$.

\begin{lemma}
\label{Lem:buyer-pricing}The following conditions are equivalent:

\begin{enumerate}
\item [$1)$]$(\gamma,\delta)\in\operatorname*{epi}z_{0}$.

\item[$2)$] There exist a strategy $(\alpha,\beta)\in\Phi$ such that
$(\alpha_{0},\beta_{0})=(\gamma,\delta)$ and a stopping time $\tau
\in\mathcal{T}$ such that $(\alpha_{\tau},\beta_{\tau})\in\operatorname*{epi}%
u_{\tau}$.

\item[$3)$] There is a superhedging strategy $(\alpha,\beta)\in\Phi$ for the
buyer such that $(\alpha_{0},\beta_{0})=(\gamma,\delta)$.
\end{enumerate}
\end{lemma}

\begin{proof}
$1)\Rightarrow2)$. If $(\gamma,\delta)\in\operatorname*{epi}z_{0}$, then
Algorithm~\ref{Alg:buyer-hedging1} gives a stopping time $\tau\in\mathcal{T}$
and a strategy $(\alpha,\beta)\in\Phi$ such that $(\alpha_{0},\beta
_{0})=(\gamma,\delta)$. We have $(\alpha_{\tau},\beta_{\tau})\in
\operatorname*{epi}u_{\tau}$ because, by construction, $(\alpha_{t},\beta
_{t})\in\operatorname*{epi}u_{t}$ on $\{\tau=t\}$ for each $t=0,1,\ldots,T$.
Condition~$2)$ is therefore satisfied.

$2)\Rightarrow3)$. This follows immediately because $(\alpha_{\tau}%
,\beta_{\tau})\in\operatorname*{epi}u_{\tau}$ is equivalent to the buyer's
superhedging condition~(\ref{Eq:buyer-superrepl-cond}).

$3)\Rightarrow1)$. Suppose that $(\alpha,\beta)\in\Phi$ is a superhedging
strategy for the buyer such that $(\alpha_{0},\beta_{0})=(\gamma,\delta)$.
Then there is a $\tau\in\mathcal{T}$ such that $(\alpha_{\tau},\beta_{\tau
})\in\operatorname*{epi}u_{\tau}$. We claim that $(\alpha_{t},\beta_{t}%
)\in\operatorname*{epi}z_{t}$ on $\{t\leq\tau\}$ for all $t=0,1,\ldots,T$,
which implies $1)$ immediately. The claim can be verified by backward
induction on~$t$. We clearly have $(\alpha_{T},\beta_{T})\in
\operatorname*{epi}u_{T}=\operatorname*{epi}z_{T}$ on $\{T\leq\tau
\}=\{T=\tau\}$. Now suppose that the claim is valid for some $t=1,\ldots,T$.
We consider two cases:

\begin{itemize}
\item  On $\{t-1=\tau\}$ we have $(\alpha_{t-1},\beta_{t-1})\in
\operatorname*{epi}u_{t-1}\subset\operatorname*{epi}z_{t-1}$
by~(\ref{Eq:Alg_buyer_min}).

\item  On $\{t\leq\tau\}$ we have $(\alpha_{t},\beta_{t})\in
\operatorname*{epi}z_{t}$ by the induction hypothesis. Because $(\alpha
_{t},\beta_{t})$ is $\mathcal{F}_{t-1}$-measurable, it follows
by~(\ref{Eq:Alg_buyer_max}) that $(\alpha_{t},\beta_{t})\in\operatorname*{epi}%
w_{t-1}$ on $\{t\leq\tau\}$. Since the strategy is self-financing,
$(\alpha_{t-1}-\alpha_{t},\beta_{t-1}-\beta_{t})\in\operatorname*{epi}%
h_{[S_{t-1}^{\mathrm{b}},S_{t-1}^{\mathrm{a}}]}$. As a result,
by~(\ref{Eq:Alg_buyer_gr}) and~(\ref{Eq:Alg_buyer_min}) we obtain%
\[
(\alpha_{t-1},\beta_{t-1})\in\operatorname*{epi}h_{[S_{t-1}^{\mathrm{b}%
},S_{t-1}^{\mathrm{a}}]}+\operatorname*{epi}w_{t-1}=\operatorname*{epi}%
v_{t-1}\subset\operatorname*{epi}z_{t-1}\quad\text{on }\{t\leq\tau\}.
\]
\end{itemize}

\noindent It follows that $(\alpha_{t-1},\beta_{t-1})\in\operatorname*{epi}%
z_{t-1}$ on $\{t-1\leq\tau\}=\{t-1=\tau\}\cup\{t\leq\tau\}$, which completes
the proof of the claim.
\end{proof}

\subsubsection{Buyer's Stopping Time and Approximate
Martingale\label{Sect:buyer-stopping-approx-mart}}

In this case there is no need for a separate algorithm. The construction of
the stopping time is already covered by the buyer's hedging
Algorithm~\ref{Alg:buyer-hedging1}, whereas the approximate martingale can be
constructed using the seller's Algorithm~\ref{Alg:opt-stop} as detailed below.

Let $\check{\tau}\in\mathcal{T}$ be the stopping time and $(\check{\alpha
},\check{\beta})\in\Phi$ the strategy constructed in
Algorithm~\ref{Alg:buyer-hedging1} starting from the portfolio $(\check
{\alpha}_{0},\check{\beta}_{0})=(z_{0}(0),0)\in\operatorname*{epi}z_{0}$,
with~$z_{0}$ from Algorithm~\ref{Alg:buyer-price}. Consider a payoff process
$(\xi^{\prime},\zeta^{\prime})$ such that for each $t=0,1,\ldots,T$%
\[
(\xi_{t}^{\prime},\zeta_{t}^{\prime})=\left\{
\begin{array}
[c]{ll}%
(-\xi_{\check{\tau}},-\zeta_{\check{\tau}}) & \text{if }t=\check{\tau},\\
(-\infty,-\infty) & \text{if }t\not =\check{\tau}.
\end{array}
\right.
\]
The mixed stopping time in the seller's Algorithm~\ref{Alg:opt-stop} for the
option~$(\xi^{\prime},\zeta^{\prime})$ can be constructed in such a way that
it takes zero values at all nodes where $t\neq\check{\tau}$, at which
$(\xi_{t}^{\prime},\zeta_{t}^{\prime})=(-\infty,-\infty)$. This mixed stopping
time must therefore be equal to~$\check{\tau}$. Algorithm~\ref{Alg:opt-stop}
also provides an approximate martingale $(\check{P},\check{S})\in
\mathcal{\bar{P}}(\check{\tau})$.

In Theorem~\ref{Thm:buyer-martingale-representation} the stopping
time~$\check{\tau}$ and approximate martingale~$(\check{P},\check{S})$ will be
related to the bid (buyer's) option price $\pi^{\mathrm{b}}(\xi,\zeta)$. For
now, we prove the following lemma.

\begin{lemma}
\label{Lem:buyer-stop-appr-mart}If $\check{\tau}\in\mathcal{T}$ is the
stopping time and $(\check{P},\check{S})\in\mathcal{\bar{P}}(\check{\tau})$
constructed as above for the buyer of an American option $(\xi,\zeta)$, then%
\[
\mathbb{E}_{\check{P}}(\xi_{\check{\tau}}+\check{S}_{\check{\tau}}%
\zeta_{\check{\tau}})=\min_{(P,S)\in\mathcal{\bar{P}}(\check{\tau})}%
\mathbb{E}_{P}(\xi_{\check{\tau}}+S_{\check{\tau}}\zeta_{\check{\tau}}%
)\geq-z_{0}(0),
\]
where $z_{0}\in\Theta$ is constructed in the buyer's pricing
Algorithm~\ref{Alg:buyer-price}.
\end{lemma}

\begin{proof}
By Theorem~\ref{Thm:seller-martingale-representation},%
\[
\pi^{\mathrm{a}}(\xi^{\prime},\zeta^{\prime})=\mathbb{E}_{\check{P}}%
(\xi_{\check{\tau}}^{\prime}+\check{S}_{\check{\tau}}\zeta_{\check{\tau}%
}^{\prime})=\max_{(P,S)\in\mathcal{\bar{P}}(\check{\tau})}\mathbb{E}_{P}%
(\xi_{\check{\tau}}^{\prime}+S_{\check{\tau}}\zeta_{\check{\tau}}^{\prime}).
\]
Since $(\xi_{\check{\tau}}^{\prime},\zeta_{\check{\tau}}^{\prime}%
)=(-\xi_{\check{\tau}},-\zeta_{\check{\tau}})$ and $\vartheta_{\check{\tau}%
}(\check{\alpha}_{\check{\tau}}+\xi_{\check{\tau}},\check{\beta}_{\check{\tau
}}+\zeta_{\check{\tau}})\geq0$, it follows by~(\ref{Eq:def-pi-a}) that%
\begin{align*}
\mathbb{E}_{\check{P}}(\xi_{\check{\tau}}+\check{S}_{\check{\tau}}%
\zeta_{\check{\tau}})  &  =\min_{(P,S)\in\mathcal{\bar{P}}(\check{\tau}%
)}\mathbb{E}_{P}(\xi_{\check{\tau}}+S_{\check{\tau}}\zeta_{\check{\tau}}%
)=-\pi^{\mathrm{a}}(\xi^{\prime},\zeta^{\prime})\\
&  =\max\left\{  \vartheta_{0}(-\alpha_{0},-\beta_{0})\,|\,(\alpha,\beta
)\in\Phi,\vartheta_{\check{\tau}}(\alpha_{\check{\tau}}+\xi_{\check{\tau}%
},\beta_{\check{\tau}}+\zeta_{\check{\tau}})\geq0\right\} \\
&  \geq\vartheta_{0}(-\check{\alpha}_{0},-\check{\beta}_{0})=-z_{0}(0),
\end{align*}
as claimed.
\end{proof}

\subsubsection{Representations of Buyer's Price}

The following result provides representations of the bid option price based on
the constructions put forward in the buyer's case. Note the appearance of pure
stopping times rather than mixed ones, which should be contrasted with
Theorem~\ref{Thm:seller-martingale-representation}.

\begin{theorem}
\label{Thm:buyer-martingale-representation}The bid (buyer's) price of an
American option with payoff process $(\xi,\zeta)$ can be represented as
follows:%
\begin{align*}
\pi^{\mathrm{b}}(\xi,\zeta)  &  =-z_{0}(0)=\mathbb{E}_{\check{P}}(\xi
_{\check{\tau}}+\check{S}_{\check{\tau}}\zeta_{\check{\tau}})=\min
_{(P,S)\in\mathcal{\bar{P}}(\check{\tau})}\mathbb{E}_{P}(\xi_{\check{\tau}%
}+S_{\check{\tau}}\zeta_{\check{\tau}})\\
&  =\max_{\tau\in\mathcal{T}}\min_{(P,S)\in\mathcal{\bar{P}}(\tau)}%
\mathbb{E}_{P}(\xi_{\tau}+S_{\tau}\zeta_{\tau})=\max_{\tau\in\mathcal{T}}%
\inf_{(P,S)\in\mathcal{P}(\tau)}\mathbb{E}_{P}(\xi_{\tau}+S_{\tau}\zeta_{\tau
}),
\end{align*}
where $z_{0}\in\Theta$ is constructed in Algorithm~\ref{Alg:buyer-price}, and
$\check{\tau}\in\mathcal{T}$ with $(\check{P},\check{S})\in\mathcal{\bar{P}%
}(\check{\tau})$ in Section~\ref{Sect:buyer-stopping-approx-mart}.
\end{theorem}

\begin{proof}
From the definition~(\ref{Eq:def-pi-b}) of $\pi^{\mathrm{b}}(\xi,\zeta)$ and
Lemma~\ref{Lem:buyer-pricing} we have%
\[
\pi^{\mathrm{b}}(\xi,\zeta)=\max\left\{  \vartheta_{0}(-\gamma,-\delta
)|(\gamma,\delta)\in\operatorname*{epi}z_{0}\right\}  .
\]
By the construction in Algorithm~\ref{Alg:buyer-price}, we have
$\operatorname{gr}_{[S_{0}^{\mathrm{b}},S_{0}^{\mathrm{a}}]}(z_{0})=z_{0}%
$.\ This means that if $\gamma\geq z_{0}(\delta)$, then $-\vartheta
_{0}(-\gamma,-\delta)=\gamma-\delta^{-}S_{0}^{\mathrm{b}}+\delta^{+}%
S_{0}^{\mathrm{a}}\geq z_{0}(\delta)-\delta^{-}S_{0}^{\mathrm{b}}+\delta
^{+}S_{0}^{\mathrm{a}}\geq z_{0}(0)$, so that $\pi^{\mathrm{b}}(\xi,\zeta
)\leq-z_{0}(0)$. Then, by Lemma~\ref{Lem:buyer-stop-appr-mart},%
\begin{equation}
\pi^{\mathrm{b}}(\xi,\zeta)\leq-z_{0}(0)\leq\mathbb{E}_{\check{P}}(\xi
_{\check{\tau}}+\check{S}_{\check{\tau}}\zeta_{\check{\tau}})=\min
_{(P,S)\in\mathcal{\bar{P}}(\check{\tau})}\mathbb{E}_{P}(\xi_{\check{\tau}%
}+S_{\check{\tau}}\zeta_{\check{\tau}}). \label{Eq:proof-buyer-repr-thm1}%
\end{equation}
Now take any $\tau\in\mathcal{T}$ and a payoff process $(\xi^{\prime}%
,\zeta^{\prime})$ such that for each $t=0,1,\ldots,T$%
\[
(\xi_{t}^{\prime},\zeta_{t}^{\prime})=\left\{
\begin{array}
[c]{ll}%
(-\xi_{\tau},-\zeta_{\tau}) & \text{if }t=\tau,\\
(-\infty,-\infty) & \text{if }t\not =\tau.
\end{array}
\right.
\]
The mixed stopping time in the seller's Algorithm~\ref{Alg:opt-stop} for the
option~$(\xi^{\prime},\zeta^{\prime})$ can be constructed in such a way that
it takes zero values at all nodes where $t\neq\tau$, at which $(\xi
_{t}^{\prime},\zeta_{t}^{\prime})=(-\infty,-\infty)$. This mixed stopping time
must therefore be equal to~$\tau$. Using
Theorem~\ref{Thm:seller-martingale-representation} and~(\ref{Eq:def-pi-a}) we
therefore find that%
\begin{multline*}
\min_{(P,S)\in\mathcal{\bar{P}}(\tau)}\mathbb{E}_{P}(\xi_{\tau}+S_{\tau}%
\zeta_{\tau})=-\max_{(P,S)\in\mathcal{\bar{P}}(\tau)}\mathbb{E}_{P}(\xi_{\tau
}^{\prime}+S_{\tau}\zeta_{\tau}^{\prime})=-\pi^{\mathrm{a}}(\xi^{\prime}%
,\zeta^{\prime})\\
=\max\left\{  \vartheta_{0}(-\alpha_{0},-\beta_{0})\,|\,(\alpha,\beta)\in
\Phi,\vartheta_{\tau}(\alpha_{\tau}+\xi_{\tau},\beta_{\tau}+\zeta_{\tau}%
)\geq0\right\}  .
\end{multline*}
It follows by (\ref{Eq:def-pi-b}) that%
\begin{align*}
\pi^{\mathrm{b}}(\xi,\zeta)  &  =\max_{\tau\in\mathcal{T}}\max\left\{
\vartheta_{0}(-\alpha_{0},-\beta_{0})|(\alpha,\beta)\in\Phi,\vartheta_{\tau
}(\alpha_{\tau}+\xi_{\tau},\beta_{\tau}+\zeta_{\tau})\geq0\right\} \\
&  =\max_{\tau\in\mathcal{T}}\min_{(P,S)\in\mathcal{\bar{P}}(\tau)}%
\mathbb{E}_{P}(\xi_{\tau}+S_{\tau}\zeta_{\tau})=\max_{\tau\in\mathcal{T}}%
\inf_{(P,S)\in\mathcal{P}(\tau)}\mathbb{E}_{P}(\xi_{\tau}+S_{\tau}\zeta_{\tau
}).
\end{align*}
The last equality is valid by Proposition~\ref{Prop:sup-max} since
$\mathcal{P}(\tau)\subset\mathcal{\bar{P}}(\tau)$. Combined with
(\ref{Eq:proof-buyer-repr-thm1}), this completes the proof because
$\check{\tau}\in\mathcal{T}$ and $(\check{P},\check{S})\in\mathcal{\bar{P}%
}(\check{\tau})$.
\end{proof}

\begin{corollary}
The self-financing strategy $(\check{\alpha},\check{\beta})\in\Phi$ and
stopping time $\check{\tau}\in\mathcal{T}$ constructed in
Algorithm~\ref{Alg:buyer-hedging1} starting from the portfolio $(\check
{\alpha}_{0},\check{\beta}_{0})=(-\pi^{\mathrm{b}}(\xi,\zeta),0)$ are optimal
for the option buyer, that is, they realise the maximum in the
definition~$(\ref{Eq:def-pi-b})$ of $\pi^{\mathrm{b}}(\xi,\zeta)$.
\end{corollary}

\section{Example\label{Sect:Example}}

\begin{example}
\label{Exl:clinical_example_new}\upshape Consider a two-step binomial tree
model with risk-free rate equal to~$0$ (all bond prices equal to~$1$) and ask
and bid stock prices $S^{\mathrm{a}},S^{\mathrm{b}}$, together with an
American option with payoff process $(\xi,\zeta)$ as in the following diagram:%
\[%
\begin{array}
[c]{ll}%
S_{0}^{\mathrm{a}}=10 & \xi_{0}=0\\
S_{0}^{\mathrm{b}}=10 & \zeta_{0}=0
\end{array}%
\begin{array}
[c]{l}%
\nearrow\\
\searrow
\end{array}%
\begin{array}
[c]{ll}%
S_{1}^{\mathrm{a}}=16 & \xi_{1}=3\\
S_{1}^{\mathrm{b}}=8 & \zeta_{1}=0\\
& \\
S_{1}^{\mathrm{a}}=6 & \xi_{1}=0\\
S_{1}^{\mathrm{b}}=6 & \zeta_{1}=0
\end{array}%
\begin{array}
[c]{l}%
\nearrow\\
\searrow\\
\\
\nearrow\\
\searrow
\end{array}%
\begin{array}
[c]{ll}%
S_{1}^{\mathrm{a}}=16 & \xi_{2}=9\\
S_{1}^{\mathrm{b}}=16 & \zeta_{2}=0\\
& \\
S_{1}^{\mathrm{a}}=10 & \xi_{2}=0\\
S_{1}^{\mathrm{b}}=10 & \zeta_{2}=0\\
& \\
S_{1}^{\mathrm{a}}=4 & \xi_{2}=0\\
S_{1}^{\mathrm{b}}=4 & \zeta_{2}=0
\end{array}
\]
The nodes in the tree at time~$1$ will be referred to as \textrm{u}
and~\textrm{d}, and those at time~$2$ as $\mathrm{u}$\textrm{u}, \textrm{ud},
\textrm{du} and~\textrm{dd}. The ask and bid stock prices as well as the
payoffs are taken to be the same at nodes \textrm{ud} and~\textrm{du} (they
are path-independent). The option is settled in cash, that is,\ $\zeta\equiv0$.

In Figure~\ref{Fig:seller_alg1} we present the construction in
Algorithm~\ref{Alg:seller-price1} for two nodes, \textrm{u} and the root node,
which are the interesting ones in this example. The construction at any of the
remaining nodes is straightforward. Looking at function~$z_{0}$, we find the
ask (seller's) price of the option to be%
\[
\pi^{\mathrm{a}}(\xi,\zeta)=z_{0}(0)=\textstyle4\frac{1}{2}.
\]
The seller's optimal strategy $(\hat{\alpha},\hat{\beta})\in\Phi$ can be
constructed by following Algorithm~\ref{Alg:seller-hedging}. In this way we
obtain%
\[%
\begin{array}
[c]{c}%
(\hat{\alpha}_{0},\hat{\beta}_{0})=\textstyle(4\frac{1}{2},0)
\end{array}%
\begin{array}
[c]{c}%
\rightarrow
\end{array}%
\begin{array}
[c]{c}%
(\hat{\alpha}_{1},\hat{\beta}_{1})=\textstyle(-3,\frac{3}{4})
\end{array}%
\begin{array}
[c]{c}%
\nearrow\\
\searrow
\end{array}%
\begin{array}
[c]{l}%
(\hat{\alpha}_{2}^{\mathrm{u}},\hat{\beta}_{2}^{\mathrm{u}})=\textstyle
(-3,\frac{3}{4})\\
\\
(\hat{\alpha}_{2}^{\mathrm{d}},\hat{\beta}_{2}^{\mathrm{d}})=\textstyle
(\frac{3}{2},0)
\end{array}
\]%
\begin{figure}
[h]
\begin{center}
\includegraphics[
height=5.4224in,
width=4.1027in
]%
{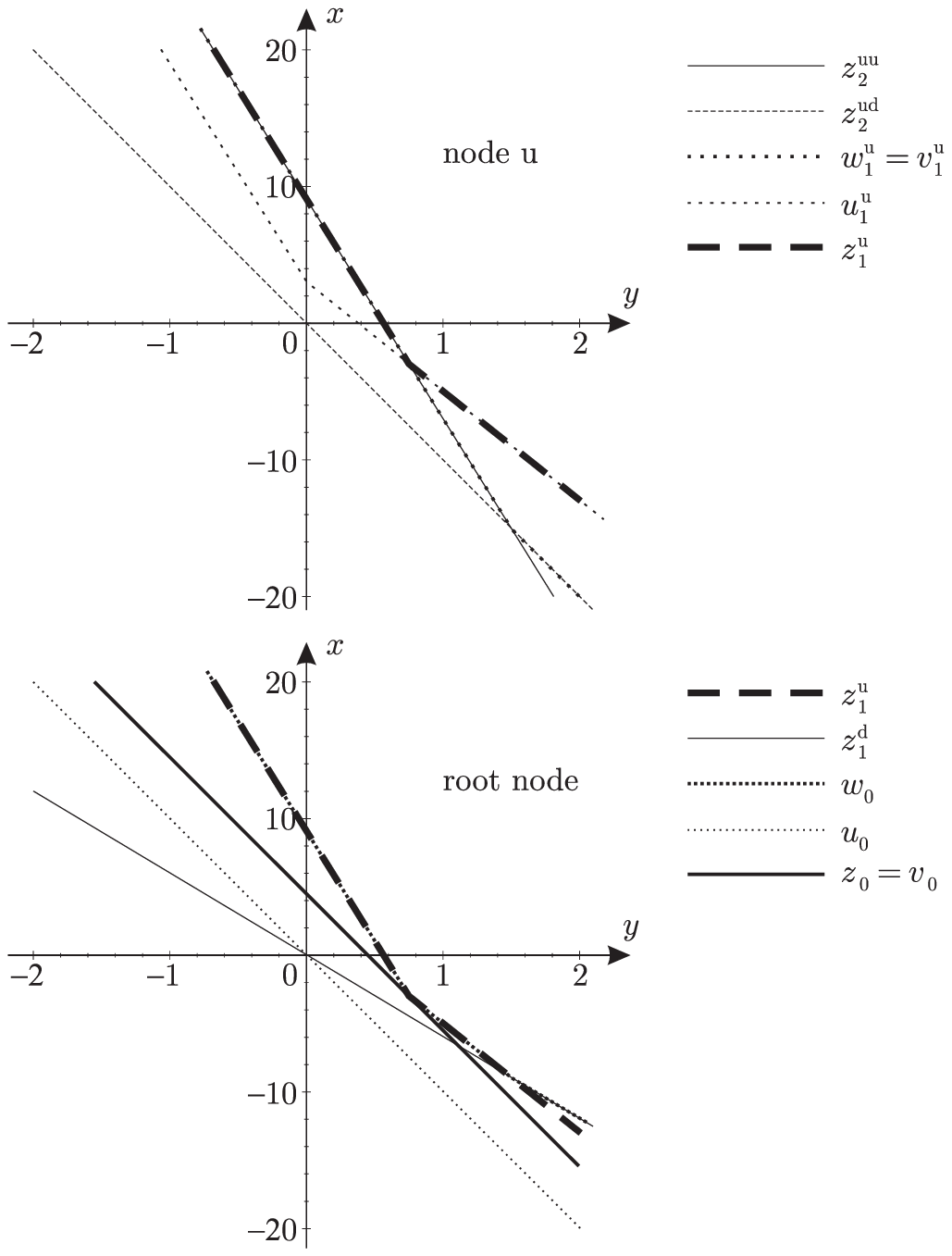}%
\caption{Algorithm~\ref{Alg:seller-price1} for the option seller in
Example~\ref{Exl:clinical_example_new} at node $\mathrm{u}$ and the root node}%
\label{Fig:seller_alg1}%
\end{center}
\end{figure}

The construction in Algorithm~\ref{Alg:seller-price2}, equivalent by convex
duality to Algorithm~\ref{Alg:seller-price1}, is presented in
Figure~\ref{Fig:seller_alg2}, also at node~\textrm{u} and the root node. By
examining the function~$Z_{0}=z_{0}^{\ast}$ (which happens to have only one
finite value in this example) we can also see that%
\[
\pi^{\mathrm{a}}(\xi,\zeta)=\max_{x\in\mathbb{R}}Z_{0}(x)=Z_{0}(10)=\textstyle
4\frac{1}{2}.
\]
A mixed stopping time~$\hat{\chi}\in\mathcal{X}$ and approximate martingale
$(\hat{P},\hat{S})\in\mathcal{\bar{P}}(\hat{\chi})$ realising the seller's
price%
\[
\pi^{\mathrm{a}}(\xi,\zeta)=\mathbb{E}_{\hat{P}}((\xi+\hat{S}\zeta)_{\hat
{\chi}})=\textstyle4\frac{1}{2}%
\]
can be constructed as in Algorithm~\ref{Alg:opt-stop}:%
\[%
\begin{array}
[c]{l}%
\hat{\chi}_{0}=0\\
\hat{P}_{0}=1\\
\hat{S}_{0}=10
\end{array}%
\begin{array}
[c]{l}%
\nearrow\\
\\
\\
\\
\searrow
\end{array}%
\begin{array}
[c]{l}%
\hat{\chi}_{1}^{\mathrm{u}}=\frac{3}{4}\\
\hat{P}_{1}^{\mathrm{u}}=1\\
\hat{S}_{1}^{\mathrm{u}}=8\\
\\
\\
\\
\\
\\
\hat{\chi}_{1}^{\mathrm{d}}=0\\
\hat{P}_{1}^{\mathrm{d}}=0\\
\hat{S}_{1}^{\mathrm{d}}=6
\end{array}%
\begin{array}
[c]{l}%
\nearrow\\
\searrow\\
\\
\\
\\
\\
\\
\\
\nearrow\\
\searrow
\end{array}%
\begin{array}
[c]{l}%
\hat{\chi}_{2}^{\mathrm{uu}}=\frac{1}{4}\\
\hat{P}_{2}^{\mathrm{uu}}=1\\
\hat{S}_{2}^{\mathrm{uu}}=16\\
\\
\hat{\chi}_{2}^{\mathrm{ud}}=\frac{1}{4}\\
\hat{P}_{2}^{\mathrm{ud}}=0\\
\hat{S}_{2}^{\mathrm{ud}}=10\\
\\
\hat{\chi}_{2}^{\mathrm{du}}=1\\
\hat{P}_{2}^{\mathrm{du}}=1\\
\hat{S}_{2}^{\mathrm{du}}=10\\
\\
\hat{\chi}_{2}^{\mathrm{dd}}=1\\
\hat{P}_{2}^{\mathrm{dd}}=0\\
\hat{S}_{2}^{\mathrm{dd}}=4
\end{array}
\]
This construction is also illustrated in Figure~\ref{Fig:seller_alg2}, which
shows the values of the processes $\hat{S},\hat{X},\hat{Y},\hat{Z},\hat
{U},\hat{V},\hat{W}$ at \textrm{u} and the root node. The values $\hat{\chi
}_{0}=0$ and $\hat{\chi}_{1}^{\mathrm{u}}=\frac{3}{4}$ can be traced back to
the following relationships, which can be seen in Figure~\ref{Fig:seller_alg2}%
:%
\[%
\begin{array}
[c]{cc}%
\hat{Y}_{0}=0\hat{S}_{0}+1\hat{X}_{0}, & \hat{Y}_{1}^{\mathrm{u}}=\frac{3}%
{4}\hat{S}_{1}^{\mathrm{u}}+\frac{1}{4}\hat{X}_{1}^{\mathrm{u}},\\
\hat{Z}_{0}=0\hat{U}_{0}+1\hat{V}_{0}, & \hat{Z}_{1}^{\mathrm{u}}=\frac{3}%
{4}\hat{U}_{1}^{\mathrm{u}}+\frac{1}{4}\hat{V}_{1}^{\mathrm{u}}.
\end{array}
\]%
\begin{figure}
[h]
\begin{center}
\includegraphics[
height=3.8069in,
width=4.4702in
]%
{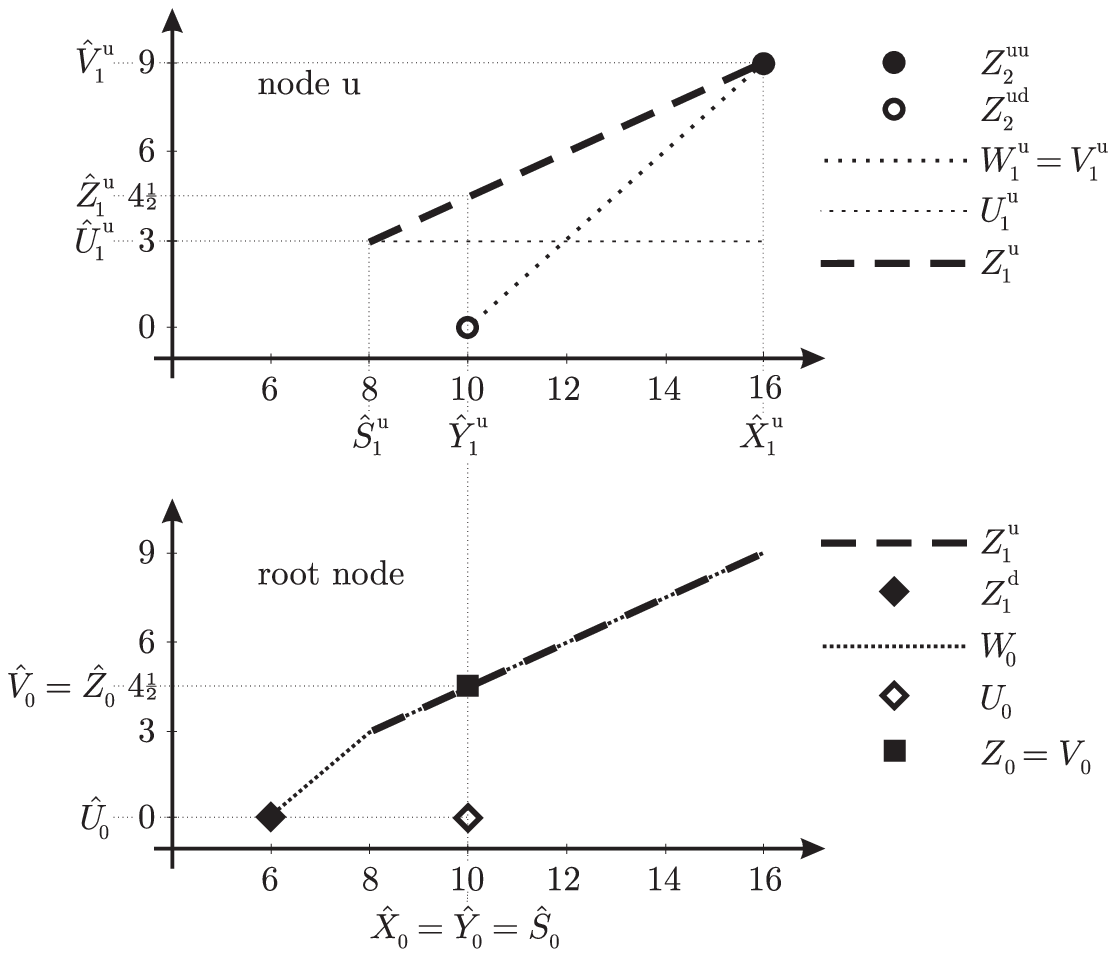}%
\caption{Algorithm~\ref{Alg:seller-price2} for the option seller in
Example~\ref{Exl:clinical_example_new} at node $\mathrm{u}$ and the root node}%
\label{Fig:seller_alg2}%
\end{center}
\end{figure}

This example also demonstrates that mixed stopping times play an essential
role in the representation of the seller's price $\pi^{\mathrm{a}}(\xi,\zeta)$
in Theorem~\ref{Thm:seller-martingale-representation}, and cannot be replaced
by pure stopping times. Indeed,%
\[
\max_{\tau\in\mathcal{T}}\max_{(P,S)\in\mathcal{\bar{P}}}\mathbb{E}_{P}%
(\xi_{\tau}+S_{\tau}\zeta_{\tau})=\textstyle
3\frac{3}{5},
\]
attained for $\tau\equiv2$, is lower than the seller's price%
\[
\pi^{\mathrm{a}}(\xi,\zeta)=\max_{\chi\in\mathcal{X}}\max_{(P,S)\in
\mathcal{\bar{P}}(\chi)}\mathbb{E}_{P}((\xi+S\zeta)_{\chi})=\textstyle
4\frac{1}{2}.
\]

The buyer's case, based on Algorithm~\ref{Alg:buyer-price}, is shown in
Figure~\ref{Fig:buyer_alg}. It involves non-convex functions such as
$z_{1}^{\mathrm{u}}$ or $w_{0}$, hence there is no convex dual counterpart.
The bid (buyer's) price is%
\[
\pi^{\mathrm{b}}(\xi,\zeta)=-z_{0}(0)=\textstyle
1\frac{1}{5}.
\]
The buyer's optimal superhedging strategy $(\check{\alpha},\check{\beta}%
)\in\Phi$ and optimal stopping time $\check{\tau}\in\mathcal{T}$ are
constructed following Algorithm~\ref{Alg:buyer-hedging1}:%
\begin{gather*}%
\begin{array}
[c]{c}%
(\check{\alpha}_{0},\check{\beta}_{0})=\textstyle(-1\frac{1}{5},0)
\end{array}%
\begin{array}
[c]{c}%
\rightarrow
\end{array}%
\begin{array}
[c]{c}%
(\check{\alpha}_{1},\check{\beta}_{1})=\textstyle(1\frac{4}{5},-\frac{3}{10})
\end{array}%
\begin{array}
[c]{c}%
\nearrow\\
\searrow
\end{array}%
\begin{array}
[c]{l}%
(\check{\alpha}_{2}^{\mathrm{u}},\check{\beta}_{2}^{\mathrm{u}})=\textstyle
(1\frac{4}{5},-\frac{3}{10})\\
\\
(\check{\alpha}_{2}^{\mathrm{d}},\check{\beta}_{2}^{\mathrm{d}})=\textstyle
(1\frac{4}{5},-\frac{3}{10})
\end{array}
\\
\check{\tau}\equiv1.
\end{gather*}
An approximate martingale $(\check{P},\check{S})\in\mathcal{\bar{P}}%
(\check{\tau})$ realising the buyer's price%
\[
\pi^{\mathrm{b}}(\xi,\zeta)=\mathbb{E}_{\check{P}}(\xi_{\check{\tau}}%
+\check{S}_{\check{\tau}}\zeta_{\check{\tau}})=\min_{(P,S)\in\mathcal{\bar{P}%
}(\check{\tau})}\mathbb{E}_{P}(\xi_{\check{\tau}}+S_{\check{\tau}}%
\zeta_{\check{\tau}})
\]
can be computed by applying Algorithm~\ref{Alg:opt-stop} to the option with
payoff $(\xi_{t}^{\prime},\zeta_{t}^{\prime})=(-\xi_{t},-\zeta_{t})$ when
$t=\check{\tau}$ and $(\xi_{t}^{\prime},\zeta_{t}^{\prime})=(-\infty,-\infty)$
when $t\neq\check{\tau}$ as explained in
Section~\ref{Sect:buyer-stopping-approx-mart}:%
\[%
\begin{array}
[c]{l}%
\check{P}_{0}=1\\
\check{S}_{0}=10
\end{array}%
\begin{array}
[c]{l}%
\nearrow\\
\\
\\
\searrow
\end{array}%
\begin{array}
[c]{l}%
\check{P}_{1}^{\mathrm{u}}=\frac{2}{5}\\
\check{S}_{1}^{\mathrm{u}}=16\\
\\
\\
\\
\\
\check{P}_{1}^{\mathrm{d}}=\frac{3}{5}\\
\check{S}_{1}^{\mathrm{d}}=6
\end{array}%
\begin{array}
[c]{l}%
\nearrow\\
\searrow\\
\\
\\
\\
\\
\nearrow\\
\searrow
\end{array}%
\begin{array}
[c]{l}%
\check{P}_{2}^{\mathrm{uu}}=\frac{2}{5}\\
\check{S}_{2}^{\mathrm{uu}}=16\\
\\
\check{P}_{2}^{\mathrm{ud}}=0\\
\check{S}_{2}^{\mathrm{ud}}=10\\
\\
\check{P}_{2}^{\mathrm{du}}=\frac{3}{5}\\
\check{S}_{2}^{\mathrm{du}}=10\\
\\
\check{P}_{2}^{\mathrm{dd}}=0\\
\check{S}_{2}^{\mathrm{dd}}=4
\end{array}
\]%
\begin{figure}
[h]
\begin{center}
\includegraphics[
height=5.4224in,
width=4.0465in
]%
{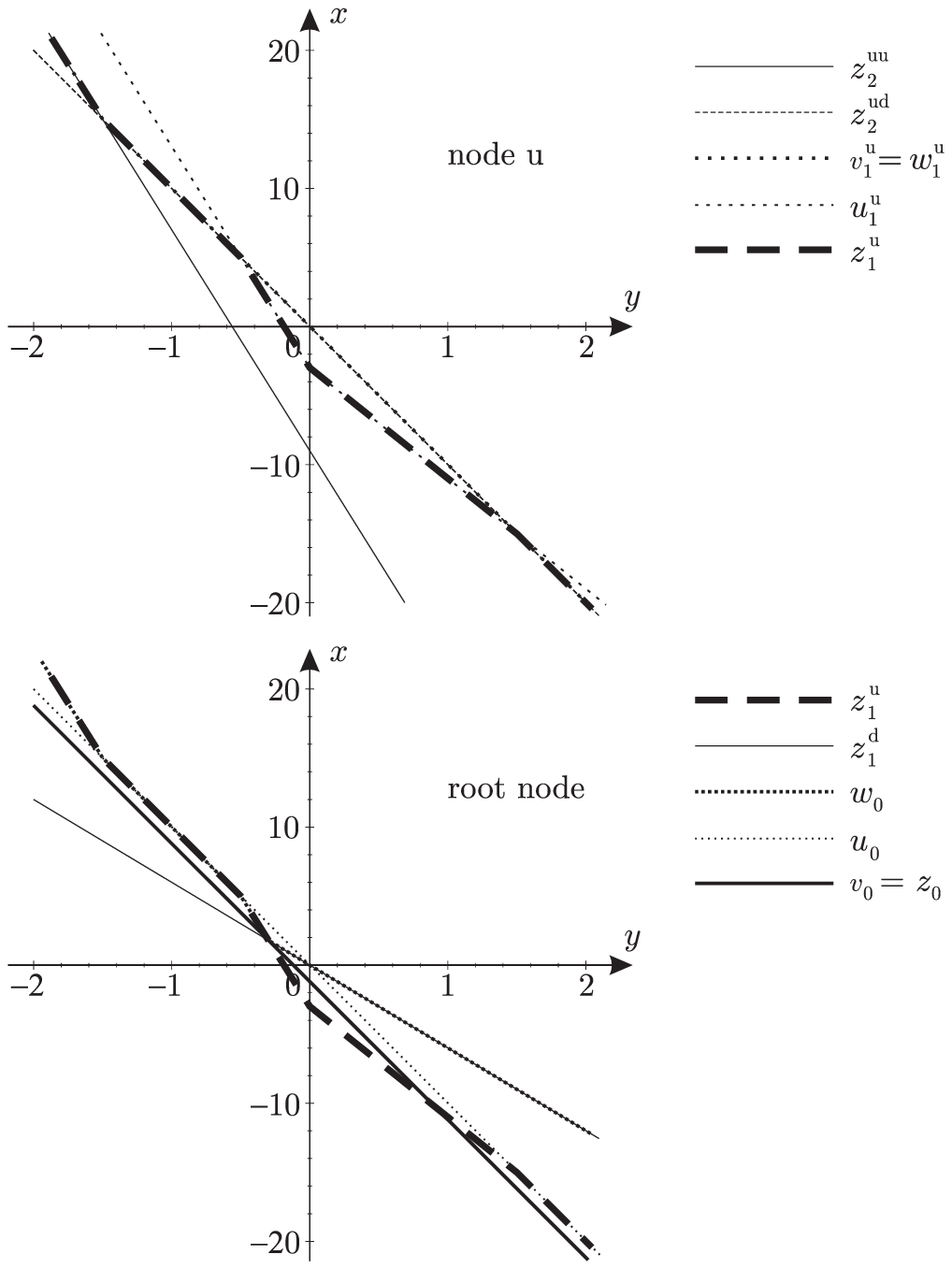}%
\caption{Algorithm~\ref{Alg:buyer-price} for the option buyer in
Example~\ref{Exl:clinical_example_new} at node $\mathrm{u}$ and the root node}%
\label{Fig:buyer_alg}%
\end{center}
\end{figure}
\end{example}

\section{Numerical Results\label{Sect:num-results}}

In this section we extend the latest published numerical results
in~\cite{PerLef04} for American puts under transaction costs very considerably
and in various directions. The algorithms proposed in the present paper can
handle not only American puts, but also arbitrary payoffs including option
baskets, cover the full range of transaction costs, and are by no means
restricted to the binomial model. The efficiency of the algorithms is
reflected in the number of time steps in the numerical examples, larger than
in \cite{PerLef04} by more than an order of magnitude. This is made possible
by the fact that in a recombinant model the computations in
Algorithms~\ref{Alg:seller-price1} and~\ref{Alg:buyer-price} grow only
polynomially with the number of time steps.

\begin{example}
\label{Exl:Am_put_after_PL04}\upshape This example is based on the setup in
Perrakis and Lefoll~\cite{PerLef04} for an American put option in a binomial
tree model under transaction costs. We reproduce the results
in~\cite{PerLef04}, and extend them to parameter ranges for which the small
transaction costs assumption imposed in~\cite{PerLef04} (namely, condition
(11) in that paper) is no longer satisfied. Thanks to the efficiency of
Algorithms~\ref{Alg:seller-price1} and~\ref{Alg:buyer-price}, we can cover a
substantially larger number of time steps and larger transaction costs than
in~\cite{PerLef04}.

The stock price process~$S$ in the binomial tree model is assumed to satisfy%
\[
S_{t}=\varepsilon_{t}S_{t-1}%
\]
for $t=1,\ldots,N$, with initial stock price $S_{0}=100$, and $\varepsilon
_{1},\varepsilon_{2},\ldots$ a sequence of independent identically distributed
random variables, each taking two possible values $e^{-\sigma\sqrt{T/N}}$ or
$e^{\sigma\sqrt{T/N}}$ with positive probability, where $\sigma=0.2$ is the
stock volatility, $T=0.25$ (three months), and $N$ is the number of time
steps. We assume a continuously compounded interest rate of~$10\%$, and a
transaction cost rate $k\in\lbrack0,1)$ so that, for $t=1,\ldots,N$ the bid
and ask stock prices are
\[
S_{t}^{\mathrm{b}}=(1-k)S_{t},\quad S_{t}^{\mathrm{a}}=(1+k)S_{t}.
\]
In line with~\cite{PerLef04}, we also assume that no transaction costs apply
at time~$0$,%
\[
S_{0}^{\mathrm{b}}=S_{0}^{\mathrm{a}}=S_{0}.
\]

We compute the ask (seller's) and bid (buyer's) prices of an American put
option exercised by the physical delivery of a portfolio $(K,-1)$ of cash and
stock, with strike price $K=100$ and time to expiry $T=0.25$. As
in~\cite{PerLef04}, the possibility that the option may never be exercised is
included. Technically, in Algorithms~\ref{Alg:seller-price1}
and~\ref{Alg:buyer-price} this is achieved by adding an extra time instant
$N+1$ to the model and setting the option payoff to be $(0,0)$ at that time.
For $N=20$ or~$40$ time steps and transaction costs $k=0.5\%$ the results in
Table~\ref{Tab:put-binom} agree with those in Table~1 in~\cite{PerLef04}. The
other results extend those in~\cite{PerLef04}.%

\begin{table}[h] \centering
$%
\begin{tabular}
[c]{c|cccccc}%
& \multicolumn{6}{|c}{$\text{Number of time steps }(N)$}\\
& 20 & 40 & 100 & 250 & 500 & 1000\\\hline
& \multicolumn{6}{|c}{$k=0.00\%$}\\
ask/bid & \multicolumn{1}{|r}{3.0485$^{\mathrm{\phantom{a}}}$} &
\multicolumn{1}{r}{3.0596$^{\mathrm{\phantom{ab}}}$} &
\multicolumn{1}{r}{3.0661$^{\mathrm{\phantom{ab}}}$} &
\multicolumn{1}{r}{3.0685$^{\mathrm{\phantom{ab}}}$} &
\multicolumn{1}{r}{3.0693$^{\mathrm{\phantom{ab}}}$} &
\multicolumn{1}{r}{3.0697$^{\mathrm{\phantom{ab}}}$}\\\hline
& \multicolumn{6}{|c}{$k=0.25\%$}\\
ask & \multicolumn{1}{|r}{3.4724$^{\mathrm{a}}$} &
\multicolumn{1}{r}{3.6366$^{\mathrm{a\phantom{b}}}$} &
\multicolumn{1}{r}{3.9348$^{\mathrm{a\phantom{b}}}$} &
\multicolumn{1}{r}{4.3691$^{\mathrm{a\phantom{b}}}$} &
\multicolumn{1}{r}{4.8194$^{\mathrm{a\phantom{b}}}$} &
\multicolumn{1}{r}{5.4023$^{\mathrm{a\phantom{b}}}$}\\
bid & \multicolumn{1}{|r}{2.5989$^{\mathrm{a}}$} &
\multicolumn{1}{r}{2.4074$^{\mathrm{a\phantom{b}}}$} &
\multicolumn{1}{r}{1.9688$^{\mathrm{a\phantom{b}}}$} &
\multicolumn{1}{r}{1.0772$^{\mathrm{a\phantom{b}}}$} &
\multicolumn{1}{r}{0.0961$^{\mathrm{a\phantom{b}}}$} &
\multicolumn{1}{r}{0.0319$^{\mathrm{a\phantom{b}}}$}\\\hline
& \multicolumn{6}{|c}{$k=0.50\%$}\\
ask & \multicolumn{1}{|r}{3.8674$^{\mathrm{\phantom{a}}}$} &
\multicolumn{1}{r}{4.1551$^{\mathrm{\phantom{ab}}}$} &
\multicolumn{1}{r}{4.6761$^{\mathrm{a\phantom{b}}}$} &
\multicolumn{1}{r}{5.4134$^{\mathrm{a\phantom{b}}}$} &
\multicolumn{1}{r}{6.1544$^{\mathrm{ab}}$} &
\multicolumn{1}{r}{7.0876$^{\mathrm{ab}}$}\\
bid & \multicolumn{1}{|r}{2.0917$^{\mathrm{\phantom{a}}}$} &
\multicolumn{1}{r}{1.5975$^{\mathrm{\phantom{ab}}}$} &
\multicolumn{1}{r}{0.2374$^{\mathrm{a\phantom{b}}}$} &
\multicolumn{1}{r}{0.0612$^{\mathrm{a\phantom{b}}}$} &
\multicolumn{1}{r}{0.0000$^{\mathrm{ab}}$} &
\multicolumn{1}{r}{0.0000$^{\mathrm{ab}}$}\\\hline
& \multicolumn{6}{|c}{$k=1.00\%$}\\
ask & \multicolumn{1}{|r}{4.5855$^{\mathrm{a}}$} &
\multicolumn{1}{r}{5.0695$^{\mathrm{a\phantom{b}}}$} &
\multicolumn{1}{r}{5.9309$^{\mathrm{ab}}$} &
\multicolumn{1}{r}{7.1120$^{\mathrm{ab}}$} &
\multicolumn{1}{r}{8.2668$^{\mathrm{ab}}$} &
\multicolumn{1}{r}{9.6890$^{\mathrm{ab}}$}\\
bid & \multicolumn{1}{|r}{0.6819$^{\mathrm{a}}$} &
\multicolumn{1}{r}{0.2589$^{\mathrm{a\phantom{b}}}$} &
\multicolumn{1}{r}{0.0000$^{\mathrm{ab}}$} &
\multicolumn{1}{r}{0.0000$^{\mathrm{ab}}$} &
\multicolumn{1}{r}{0.0000$^{\mathrm{ab}}$} &
\multicolumn{1}{r}{0.0000$^{\mathrm{ab}}$}\\\hline
& \multicolumn{6}{|c}{$k=2.00\%$}\\
ask & \multicolumn{1}{|r}{5.8274$^{\mathrm{a}}$} &
\multicolumn{1}{r}{6.5985$^{\mathrm{ab}}$} &
\multicolumn{1}{r}{7.9437$^{\mathrm{ab}}$} &
\multicolumn{1}{r}{9.7499$^{\mathrm{ab}}$} &
\multicolumn{1}{r}{11.4706$^{\mathrm{ab}}$} &
\multicolumn{1}{r}{13.5544$^{\mathrm{ab}}$}\\
bid & \multicolumn{1}{|r}{0.0492$^{\mathrm{a}}$} &
\multicolumn{1}{r}{0.0000$^{\mathrm{ab}}$} &
\multicolumn{1}{r}{0.0000$^{\mathrm{ab}}$} &
\multicolumn{1}{r}{0.0000$^{\mathrm{ab}}$} &
\multicolumn{1}{r}{0.0000$^{\mathrm{ab}}$} &
\multicolumn{1}{r}{0.0000$^{\mathrm{ab}}$}\\\hline
\multicolumn{7}{l}{$^{\mathrm{a}}$~{\small Not in~\cite{PerLef04}}}\\
\multicolumn{7}{l}{$^{\mathrm{b}}$~{\small Small transaction costs condition
(11) of~\cite{PerLef04} not satisfied}}%
\end{tabular}
$\caption{American put prices in the binomial model of~\cite{PerLef04}%
, Example~\ref{Exl:Am_put_after_PL04}\label{Tab:put-binom}}%
\end{table}%
\end{example}

\begin{example}
\label{Exl:Am-bull-binom}\upshape Within the same binomial model of stock
prices with transaction costs as in~\cite{PerLef04} described in
Example~\ref{Exl:Am_put_after_PL04} (with no transaction costs at
time~$0$)\ we consider an American bull spread, a basket consisting of a long
call with strike price~$95$ and a short call with strike price~$105$. Assume
that the bull spread is settled in cash, with payoff process $(S_{t}%
-95)^{+}-(S_{t}-105)^{+}$ and time to expiry $T=0.25$ (three months). The ask
and bid prices of the bull spread are presented in Table~\ref{Tab:bull-binom}.%

\begin{table}[h] \centering
$%
\begin{tabular}
[c]{c|cccccc}%
& \multicolumn{6}{|c}{$\text{Number of time steps }(N)$}\\
& 20 & 40 & 100 & 250 & 500 & 1000\\\hline
& \multicolumn{6}{|c}{$k=0.00\%$}\\
ask/bid & \multicolumn{1}{|r}{7.1688} & \multicolumn{1}{r}{7.2519} &
\multicolumn{1}{r}{7.2291} & \multicolumn{1}{r}{7.2023} &
\multicolumn{1}{r}{7.2576} & \multicolumn{1}{r}{7.2361}\\\hline
& \multicolumn{6}{|c}{$k=0.25\%$}\\
ask & \multicolumn{1}{|r}{7.4267} & \multicolumn{1}{r}{7.5672} &
\multicolumn{1}{r}{7.6538} & \multicolumn{1}{r}{7.8130} &
\multicolumn{1}{r}{8.3572} & \multicolumn{1}{r}{8.5756}\\
bid & \multicolumn{1}{|r}{6.8820} & \multicolumn{1}{r}{6.8793} &
\multicolumn{1}{r}{6.6756} & \multicolumn{1}{r}{6.3090} &
\multicolumn{1}{r}{5.9824} & \multicolumn{1}{r}{5.9202}\\\hline
& \multicolumn{6}{|c}{$k=0.50\%$}\\
ask & \multicolumn{1}{|r}{7.6616} & \multicolumn{1}{r}{7.8539} &
\multicolumn{1}{r}{8.2783} & \multicolumn{1}{r}{8.6371} &
\multicolumn{1}{r}{8.8761} & \multicolumn{1}{r}{8.9089}\\
bid & \multicolumn{1}{|r}{6.5599} & \multicolumn{1}{r}{6.4183} &
\multicolumn{1}{r}{5.8591} & \multicolumn{1}{r}{5.7264} &
\multicolumn{1}{r}{5.7124} & \multicolumn{1}{r}{5.6683}\\\hline
& \multicolumn{6}{|c}{$k=1.00\%$}\\
ask & \multicolumn{1}{|r}{8.1274} & \multicolumn{1}{r}{8.5640} &
\multicolumn{1}{r}{9.0392} & \multicolumn{1}{r}{9.1109} &
\multicolumn{1}{r}{9.2269} & \multicolumn{1}{r}{9.2415}\\
bid & \multicolumn{1}{|r}{5.7698} & \multicolumn{1}{r}{5.5778} &
\multicolumn{1}{r}{5.3979} & \multicolumn{1}{r}{5.2908} &
\multicolumn{1}{r}{5.2816} & \multicolumn{1}{r}{5.2413}\\\hline
& \multicolumn{6}{|c}{$k=2.00\%$}\\
ask & \multicolumn{1}{|r}{9.2537} & \multicolumn{1}{r}{9.4922} &
\multicolumn{1}{r}{9.5584} & \multicolumn{1}{r}{9.5733} &
\multicolumn{1}{r}{9.6343} & \multicolumn{1}{r}{9.6127}\\
bid & \multicolumn{1}{|r}{5.0000} & \multicolumn{1}{r}{5.0000} &
\multicolumn{1}{r}{5.0000} & \multicolumn{1}{r}{5.0000} &
\multicolumn{1}{r}{5.0000} & \multicolumn{1}{r}{5.0000}%
\end{tabular}
$\caption{American bull spread prices in the binomial model, Example~\ref
{Exl:Am-bull-binom}\label{Tab:bull-binom}}%
\end{table}%
\end{example}

\begin{example}
\label{Exl:Am-bull-trinom}\upshape We take the same American bull spread as in
Example~\ref{Exl:Am-bull-binom}, that is, a basket consisting of a long call
with strike~$95$ and a short call with strike~$105$, both settled in cash and
expiring at $T=0.25$ (three months), this time in the trinomial tree model
with stock prices $S_{0}=100$ and%
\[
S_{t}=\varepsilon_{t}S_{t-1}%
\]
for $t=1,\ldots,N$, where $\varepsilon_{1},\ldots,\varepsilon_{N}$ are
independent identically distributed random variables, each taking three
possible values $e^{\sigma\sqrt{T/N}}$ or $1$ or $e^{-\sigma\sqrt{T/N}}$. The
bid-ask spreads are defined in terms of the transaction cost rate $k\in
\lbrack0,1)$%
\[
S_{t}^{\mathrm{b}}=(1-k)S_{t},\quad S_{t}^{\mathrm{a}}=(1+k)S_{t}%
\]
for all $t=1,\ldots,T$. By analogy to the binomial model in
Examples~\ref{Exl:Am_put_after_PL04} and~\ref{Exl:Am-bull-binom}, we assume
that there are no transaction costs at time~$0$, so that $S_{0}^{\mathrm{a}%
}=S_{0}^{\mathrm{b}}=S_{0}$. We take $\sigma=0.2$ and a continuously
compounded interest rate of~$10\%$. The ask and bid prices for this option
computed by means of Algorithms~\ref{Alg:seller-price1}
and~\ref{Alg:buyer-price} are presented in Table~\ref{Tab:bull-trinom}.%

\begin{table}[h] \centering
$%
\begin{tabular}
[c]{c|cccccc}%
& \multicolumn{6}{|c}{$\text{Number of time steps }(N)$}\\
& 20 & 40 & 100 & 250 & 500 & 1000\\\hline
& \multicolumn{6}{|c}{$k=0.00\%$}\\
ask & \multicolumn{1}{|r}{7.4507} & \multicolumn{1}{r}{7.5825} &
\multicolumn{1}{r}{7.6954} & \multicolumn{1}{r}{7.7718} &
\multicolumn{1}{r}{7.8340} & \multicolumn{1}{r}{7.8702}\\
bid & \multicolumn{1}{|r}{6.2780} & \multicolumn{1}{r}{6.3117} &
\multicolumn{1}{r}{6.2696} & \multicolumn{1}{r}{6.2437} &
\multicolumn{1}{r}{6.2977} & \multicolumn{1}{r}{6.2859}\\\hline
& \multicolumn{6}{|c}{$k=0.25\%$}\\
ask & \multicolumn{1}{|r}{7.8012} & \multicolumn{1}{r}{8.0152} &
\multicolumn{1}{r}{8.2262} & \multicolumn{1}{r}{8.4083} &
\multicolumn{1}{r}{8.5873} & \multicolumn{1}{r}{8.6322}\\
bid & \multicolumn{1}{|r}{6.0191} & \multicolumn{1}{r}{6.0342} &
\multicolumn{1}{r}{5.9580} & \multicolumn{1}{r}{5.8900} &
\multicolumn{1}{r}{5.9054} & \multicolumn{1}{r}{5.8699}\\\hline
& \multicolumn{6}{|c}{$k=0.50\%$}\\
ask & \multicolumn{1}{|r}{8.1308} & \multicolumn{1}{r}{8.4095} &
\multicolumn{1}{r}{8.6574} & \multicolumn{1}{r}{8.7313} &
\multicolumn{1}{r}{8.8778} & \multicolumn{1}{r}{8.9090}\\
bid & \multicolumn{1}{|r}{5.7705} & \multicolumn{1}{r}{5.7751} &
\multicolumn{1}{r}{5.6739} & \multicolumn{1}{r}{5.6250} &
\multicolumn{1}{r}{5.6509} & \multicolumn{1}{r}{5.6199}\\\hline
& \multicolumn{6}{|c}{$k=1.00\%$}\\
ask & \multicolumn{1}{|r}{8.7576} & \multicolumn{1}{r}{8.9660} &
\multicolumn{1}{r}{9.0482} & \multicolumn{1}{r}{9.1110} &
\multicolumn{1}{r}{9.2282} & \multicolumn{1}{r}{9.2415}\\
bid & \multicolumn{1}{|r}{5.3123} & \multicolumn{1}{r}{5.3053} &
\multicolumn{1}{r}{5.2201} & \multicolumn{1}{r}{5.1818} &
\multicolumn{1}{r}{5.2100} & \multicolumn{1}{r}{5.1858}\\\hline
& \multicolumn{6}{|c}{$k=2.00\%$}\\
ask & \multicolumn{1}{|r}{9.3461} & \multicolumn{1}{r}{9.5141} &
\multicolumn{1}{r}{9.5657} & \multicolumn{1}{r}{9.5733} &
\multicolumn{1}{r}{9.6353} & \multicolumn{1}{r}{9.6127}\\
bid & \multicolumn{1}{|r}{5.0000} & \multicolumn{1}{r}{5.0000} &
\multicolumn{1}{r}{5.0000} & \multicolumn{1}{r}{5.0000} &
\multicolumn{1}{r}{5.0000} & \multicolumn{1}{r}{5.0000}%
\end{tabular}
$\caption{American bull spread prices in the trinomial model, Example~\ref
{Exl:Am-bull-trinom}\label{Tab:bull-trinom}}%
\end{table}%
\end{example}

\section{Appendix: Technical Results\label{Sect:Appendix}}

\begin{proposition}
\label{Prop:P-inclusions}For each $\chi\in\mathcal{X}$%
\[
\mathcal{P}\subset\mathcal{P}(\chi),\quad\mathcal{\bar{P}}\subset
\mathcal{\bar{P}}(\chi).
\]
\end{proposition}

\begin{proof}
To prove that $\mathcal{P}\subset\mathcal{P}(\chi)$, take any $(P,S)\in
\mathcal{P}$ and any $\chi\in\mathcal{X}$. Because $S_{t}^{\mathrm{b}}\leq
S_{t}\leq S_{t}^{\mathrm{a}}$, it is sufficient to show that for each
$t=0,1,\ldots,T$%
\begin{equation}
\chi_{t+1}^{\ast}S_{t}=\mathbb{E}_{P}(S_{t+1}^{\chi^{\ast}}|\mathcal{F}_{t}).
\label{Eq:Lem-P-in-Pofchi}%
\end{equation}
We proceed by backward induction. For $t=T$ both sides
of~(\ref{Eq:Lem-P-in-Pofchi}) are equal to zero. Suppose that
(\ref{Eq:Lem-P-in-Pofchi}) holds for some $t=1,\ldots,T$. Then, since
$\chi^{\ast}$~is a predictable process and $S$~is a martingale under~$P$,%
\begin{align*}
\mathbb{E}_{P}(S_{t}^{\chi^{\ast}}|\mathcal{F}_{t-1})  &  =\mathbb{E}_{P}%
(\chi_{t}S_{t}+S_{t+1}^{\chi^{\ast}}|\mathcal{F}_{t-1})=\mathbb{E}_{P}%
(\chi_{t}S_{t}+\mathbb{E}_{P}(S_{t+1}^{\chi^{\ast}}|\mathcal{F}_{t}%
)|\mathcal{F}_{t-1})\\
&  =\mathbb{E}_{P}(\chi_{t}S_{t}+\chi_{t+1}^{\ast}S_{t}|\mathcal{F}%
_{t-1})=\mathbb{E}_{P}(\chi_{t}^{\ast}S_{t}|\mathcal{F}_{t-1})=\chi_{t}^{\ast
}\mathbb{E}_{P}(S_{t}|\mathcal{F}_{t-1})\\
&  =\chi_{t}^{\ast}S_{t-1},
\end{align*}
completing the induction step. The proof that $\mathcal{\bar{P}}%
\subset\mathcal{\bar{P}}(\chi)$ is very similar.
\end{proof}

\begin{proposition}
\label{Prop:EchiPS_le_hedgingcost}Let $(\alpha,\beta)\in\Phi$ be a
superhedging strategy for the seller of an American option with payoff process
$(\xi,\zeta)$. Then for every $\chi\in\mathcal{X}$\ and every $(P,S)\in
\mathcal{\bar{P}}(\chi)$%
\[
\mathbb{E}_{P}(\left(  \xi+S\zeta\right)  _{\chi})\leq-\vartheta_{0}%
(-\alpha_{0},-\beta_{0}).
\]
\end{proposition}

\begin{proof}
The self-financing condition (\ref{Eq:self-fin}) satisfied by $\left(
\alpha,\beta\right)  $, along with inequalities~(\ref{Eq:def-P-of-chi_a}),
(\ref{Eq:def-P-of-chi_b})\ from the definition of~$\mathcal{\bar{P}}(\chi)$
imply that%
\begin{equation}
\mathbb{E}_{P}(\chi_{t+1}^{\ast}\alpha_{t}+S_{t+1}^{\chi^{\ast}}\beta
_{t}|\mathcal{F}_{t})\geq\mathbb{E}_{P}(\chi_{t+1}^{\ast}\alpha_{t+1}%
+S_{t+1}^{\chi^{\ast}}\beta_{t+1}|\mathcal{F}_{t}) \label{Eq:Lem-1a}%
\end{equation}
for each $t=0,1,\ldots,T$. We shall prove by backward induction that%
\begin{equation}
\mathbb{E}_{P}(\chi_{t}^{\ast}\alpha_{t}+S_{t}^{\chi^{\ast}}\beta
_{t}|\mathcal{F}_{t})\geq\mathbb{E}_{P}((\alpha+S\beta)_{t}^{\chi^{\ast}%
}|\mathcal{F}_{t}) \label{Eq:Lem-1c}%
\end{equation}
for each $t=0,1,\ldots,T$. Inequality (\ref{Eq:Lem-1c}) holds for $t=T$ since
both sides are equal to $\chi_{T}(\alpha_{T}+S_{T}\beta_{T})$. Suppose that
(\ref{Eq:Lem-1c}) holds for some $t=1,\ldots,T$. Then by~(\ref{Eq:Lem-1a})%
\begin{align*}
\mathbb{E}_{P}(\chi_{t-1}^{\ast}\alpha_{t-1}  &  +S_{t-1}^{\chi^{\ast}}%
\beta_{t-1}|\mathcal{F}_{t-1})\\
&  =\chi_{t-1}(\alpha_{t-1}+S_{t-1}\beta_{t-1})+\mathbb{E}_{P}(\chi_{t}^{\ast
}\alpha_{t-1}+S_{t}^{\chi^{\ast}}\beta_{t-1}|\mathcal{F}_{t-1})\\
&  \geq\chi_{t-1}(\alpha_{t-1}+S_{t-1}\beta_{t-1})+\mathbb{E}_{P}(\chi
_{t}^{\ast}\alpha_{t}+S_{t}^{\chi^{\ast}}\beta_{t}|\mathcal{F}_{t-1})\\
&  =\chi_{t-1}(\alpha_{t-1}+S_{t-1}\beta_{t-1})+\mathbb{E}_{P}(\mathbb{E}%
_{P}(\chi_{t}^{\ast}\alpha_{t}+S_{t}^{\chi^{\ast}}\beta_{t}|\mathcal{F}%
_{t})|\mathcal{F}_{t-1})\\
&  \geq\chi_{t-1}(\alpha_{t-1}+S_{t-1}\beta_{t-1})+\mathbb{E}_{P}%
(\mathbb{E}_{P}((\alpha+S\beta)_{t}^{\chi^{\ast}}|\mathcal{F}_{t}%
)|\mathcal{F}_{t-1})\\
&  =\chi_{t-1}(\alpha_{t-1}+S_{t-1}\beta_{t-1})+\mathbb{E}_{P}((\alpha
+S\beta)_{t}^{\chi^{\ast}}|\mathcal{F}_{t-1})\\
&  =\mathbb{E}_{P}((\alpha+S\beta)_{t-1}^{\chi^{\ast}}|\mathcal{F}_{t-1}),
\end{align*}
completing the proof of~(\ref{Eq:Lem-1c}). In particular, since $\chi
_{0}^{\ast}=1$ and $\mathbb{E}_{P}(S_{0}^{\chi^{\ast}})=\mathbb{E}_{P}%
(S_{\chi})$, inequality (\ref{Eq:Lem-1c}) for $t=0$ implies that%
\[
\alpha_{0}+\mathbb{E}_{P}(S_{\chi})\beta_{0}=\mathbb{E}_{P}(\chi_{0}^{\ast
}\alpha_{0}+S_{0}^{\chi^{\ast}}\beta_{0})\geq\mathbb{E}_{P}((\alpha
+S\beta)_{0}^{\chi^{\ast}})=\mathbb{E}_{P}((\alpha+S\beta)_{\chi}).
\]
Because $S_{0}^{\mathrm{b}}\leq\mathbb{E}_{P}(S_{\chi})\leq S_{0}^{\mathrm{a}%
}$, we have $-\vartheta_{0}(-\alpha_{0},-\beta_{0})\geq\alpha_{0}%
+\mathbb{E}_{P}(S_{\chi})\beta_{0}$. Since $(\alpha,\beta)$ is a superhedging
strategy for the seller,
\[
\vartheta_{t}(\alpha_{t}-\xi_{t},\beta_{t}-\zeta_{t})\geq0.
\]
This, together with the inequalities $S_{t}^{\mathrm{b}}\leq S_{t}\leq
S_{t}^{\mathrm{a}}$, gives $\alpha_{t}+S_{t}\beta_{t}\geq\xi_{t}+S_{t}%
\zeta_{t}$. It follows that $(\alpha+S\beta)_{\chi}\geq(\xi+S\zeta)_{\chi}$
for each $t=0,1,\ldots,T$. We therefore obtain%
\[
-\vartheta_{0}(-\alpha_{0},-\beta_{0})\geq\alpha_{0}+\mathbb{E}_{P}(S_{\chi
})\beta_{0}\geq\mathbb{E}_{P}((\alpha+S\beta)_{\chi})\geq\mathbb{E}_{P}%
((\xi+S\zeta)_{\chi}),
\]
as claimed.
\end{proof}

\begin{proposition}
\label{Prop:sup-max}Let $(\xi,\zeta)$ be the payoff process of an American
option. Then for any $\delta>0$, any mixed stopping time $\chi\in\mathcal{X}$
and any $(\bar{P},\bar{S})\in\mathcal{\bar{P}}(\chi)$ there exists a pair
$(P^{\delta},S^{\delta})\in\mathcal{P}(\chi)$ such that%
\begin{equation}
\left|  \mathbb{E}_{P^{\delta}}((\xi+S^{\delta}\zeta)_{\chi})-\mathbb{E}%
_{\bar{P}}((\xi+\bar{S}\zeta)_{\chi})\right|  <\delta.
\label{Eq:max-sup-delta}%
\end{equation}
\end{proposition}

\begin{proof}
Due to the lack of arbitrage, by the result of Jouini and
Kallal~\cite{JouKal95}, there exists some $(P,S)\in\mathcal{P}$. If
$\mathbb{E}_{P}((\xi+S\zeta)_{\chi})=\mathbb{E}_{\bar{P}}((\xi+\bar{S}%
\zeta)_{\chi})$, then (\ref{Eq:max-sup-delta}) is trivial because
$\mathcal{P}\subset\mathcal{P}(\chi)$. If this is not the case, take
any~$\varepsilon$ such that%
\[
0<\varepsilon<\min\left\{  1,\frac{\delta}{\left|  \mathbb{E}_{P}\left(
(\xi+S\zeta)_{\chi}\right)  -\mathbb{E}_{\bar{P}}\left(  (\xi+\bar{S}%
\zeta)_{\chi}\right)  \right|  }\right\}  ,
\]
and put%
\begin{align*}
P^{\delta}  &  =(1-\varepsilon)\bar{P}+\varepsilon P,\\
S_{t}^{\delta}  &  =\mathbb{E}_{P^{\delta}}\!\!\left(  \left.  (1-\varepsilon
)\bar{S}_{t}\frac{d\bar{P}}{dP^{\delta}}+\varepsilon S_{t}\frac{dP}%
{dP^{\delta}}\right|  \mathcal{F}_{t}\right)
\end{align*}
for each $t=0,1,\ldots,T$. It follows that $P^{\delta}$ is a probability
measure equivalent to~$Q$. It also follows that%
\[
S_{t}^{\delta}\leq S_{t}^{\mathrm{a}}\mathbb{E}_{P^{\delta}}\!\!\left(
\left.  (1-\varepsilon)\frac{d\bar{P}}{dP^{\delta}}+\varepsilon\frac
{dP}{dP^{\delta}}\right|  \mathcal{F}_{t}\right)  =S_{t}^{\mathrm{a}}%
\]
and, in a similar way, that%
\[
S_{t}^{\mathrm{b}}\leq S_{t}^{\delta}%
\]
for any $t=0,1,\ldots,T$. Next,%
\begin{align*}
&  \mathbb{E}_{P^{\delta}}((S^{\delta})_{t+1}^{\chi^{\ast}}|\mathcal{F}_{t})\\
&  \quad\quad=(1-\varepsilon)\mathbb{E}_{P^{\delta}}\!\!\left(  \left.
\bar{S}_{t+1}^{\chi^{\ast}}\frac{d\bar{P}}{dP^{\delta}}\right|  \mathcal{F}%
_{t}\right)  +\varepsilon\mathbb{E}_{P^{\delta}}\!\!\left(  \left.
S_{t+1}^{\chi^{\ast}}\frac{dP}{dP^{\delta}}\right|  \mathcal{F}_{t}\right) \\
&  \quad\quad=(1-\varepsilon)\mathbb{E}_{\bar{P}}(\bar{S}_{t+1}^{\chi^{\ast}%
}|\mathcal{F}_{t})\mathbb{E}_{P^{\delta}}\!\!\left(  \left.  \frac{d\bar{P}%
}{dP^{\delta}}\right|  \mathcal{F}_{t}\right)  +\varepsilon\mathbb{E}%
_{P}(S_{t+1}^{\chi^{\ast}}|\mathcal{F}_{t})\mathbb{E}_{P^{\delta}}\!\!\left(
\left.  \frac{dP}{dP^{\delta}}\right|  \mathcal{F}_{t}\right) \\
&  \quad\quad\leq\chi_{t+1}^{\ast}S_{t}^{\mathrm{a}}\left(  (1-\varepsilon
)\mathbb{E}_{P^{\delta}}\!\!\left(  \left.  \frac{d\bar{P}}{dP^{\delta}%
}\right|  \mathcal{F}_{t}\right)  +\varepsilon\mathbb{E}_{P^{\delta}%
}\!\!\left(  \left.  \frac{dP}{dP^{\delta}}\right|  \mathcal{F}_{t}\right)
\right)  =\chi_{t+1}^{\ast}S_{t}^{\mathrm{a}}%
\end{align*}
and, similarly,%
\[
\chi_{t+1}^{\ast}S_{t}^{\mathrm{b}}\leq\mathbb{E}_{P^{\delta}}((S^{\delta
})_{t+1}^{\chi^{\ast}}|\mathcal{F}_{t})
\]
for any $t=0,1,\ldots,T$. As a result, $(P^{\delta},S^{\delta})\in
\mathcal{P}(\chi)$. Moreover,
\begin{align*}
\mathbb{E}_{P^{\delta}}((\xi+S^{\delta}\zeta)_{\chi})  &  =\mathbb{E}%
_{P^{\delta}}(\xi_{\chi})+\mathbb{E}_{P^{\delta}}((S^{\delta}\zeta)_{\chi})\\
&  =(1-\varepsilon)\mathbb{E}_{\bar{P}}(\xi_{\chi})\mathbb{+\varepsilon E}%
_{P}(\xi_{\chi})+(1-\varepsilon)\mathbb{E}_{\bar{P}}((\bar{S}\zeta)_{\chi
})+\varepsilon\mathbb{E}_{P}((S\zeta)_{\chi})\\
&  =(1-\varepsilon)\mathbb{E}_{\bar{P}}((\xi+\bar{S}\zeta)_{\chi}%
)+\varepsilon\mathbb{E}_{P}((\xi+S\zeta)_{\chi}),
\end{align*}
which implies that%
\[
\left|  \mathbb{E}_{P^{\delta}}((\xi+S^{\delta}\zeta)_{\chi})-\mathbb{E}%
_{\bar{P}}((\xi+\bar{S}\zeta)_{\chi})\right|  =\varepsilon\left|
\mathbb{E}_{P}((\xi+S\zeta)_{\chi})-\mathbb{E}_{\bar{P}}((\xi+\bar{S}%
\zeta)_{\chi})\right|  <\delta.
\]
\end{proof}

\bibliographystyle{amsalpha}
\bibliography{TrCosts_070708}

\providecommand{\bysame}{\leavevmode\hbox to3em{\hrulefill}\thinspace}
\providecommand{\MR}{\relax\ifhmode\unskip\space\fi MR }
\providecommand{\MRhref}[2]{%
  \href{http://www.ams.org/mathscinet-getitem?mr=#1}{#2}
}
\providecommand{\href}[2]{#2}
\begin{thebibliography}{KRS03}

\bibitem[BC77]{BaxCha77}
S.~Baxter and R.~Chacon, \emph{Compactness of stopping times}, Z.~Wahrsch.
  verw. Gebiete \textbf{40} (1977), 169--181.

\bibitem[BT05]{BouTem05}
B.~Bouchard and E.~Temam, \emph{On the hedging of {American} options in
  discrete time markets with proportional transaction costs}, Electronic
  Journal of Probability \textbf{10} (2005), 746--760.

\bibitem[CJ01]{ChaJha01}
P.~Chalasani and S.~Jha, \emph{Randomized stopping times and {A}merican option
  pricing with transaction costs}, Math. Finance \textbf{1} (2001), 33--77.

\bibitem[CP04]{ConsPer04}
G.M. Constantinides and S.~Perrakis, \emph{Stochastic dominance bounds on
  {A}merican option prices in markets with frictions}, Working paper,
  University of Chicago, 2004.

\bibitem[CPS07]{ChenPalSheu05}
G.-Y. Chen, K.~Palmer, and Y.-C. Sheu, \emph{The least cost super replicating
  portfolio for short puts and calls in the {Boyle}-{Vorst} model with
  transaction costs}, Review of Quantitative Finance and Accounting (2007), to
  appear.

\bibitem[CRS71]{ChoRobSie71}
Y.S. Chow, H.~Robbins, and D.~Siegmund, \emph{Great expectations: The theory of
  optimal stopping}, Houghton Mifflin, Boston, 1971.

\bibitem[CZ01]{ConsZha01}
G.M. Constantinides and T.~Zariphopoulou, \emph{Bounds on derivative prices in
  an intertemporal setting with proportional transaction costs and multiple
  securities}, Math. Finance \textbf{11} (2001), 331--346.

\bibitem[DZ95]{DavZha95}
M.H.A. Davis and T.~Zariphopoulou, \emph{American options and transaction
  fees}, Mathematical Finance (M.H.A. Davis et~al., eds.), IMA Volumes in
  Mathematics and Its Applications, vol.~65, Springer, New York, 1995,
  pp.~47--61.

\bibitem[JK95]{JouKal95}
E.~Jouini and H.~Kallal, \emph{Martingales and arbitrage in securities markets
  with transaction costs}, J.~Econom. Theory \textbf{66} (1995), 178--197.

\bibitem[JLR03]{JakuLeveRyz03}
P.~Jakubenas, S.~Levental, and M.~Ryznar, \emph{The super-replication problem
  via probabilistic methods}, Ann. Appl. Probab. \textbf{13} (2003), 742--773.

\bibitem[Koc99]{Koc99}
M.~Koci{\'n}ski, \emph{Optimality of the replicating strategy for {A}merican
  options}, Appl. Math. (Warsaw) \textbf{26} (1999), 93--105.

\bibitem[Koc01]{Koc01}
\bysame, \emph{Pricing of the {A}merican option in discrete time with
  proportional transaction costs}, Math. Methods Oper. Res. \textbf{53} (2001),
  67--88.

\bibitem[KRS02]{KabRasStri02}
Y.~Kabanov, M.~R{\'a}sonyi, and C.~Stricker, \emph{No arbitrage criteria for
  financial markets with efficient friction}, Finance Stoch. \textbf{6} (2002),
  371--382.

\bibitem[KRS03]{KabRasStri03}
\bysame, \emph{On the closedness of sums of convex cones in~{$L^0$} and the
  robust no-arbitrage property}, Finance Stoch. \textbf{7} (2003), 403--411.

\bibitem[KS01]{KabStr01}
Y.~Kabanov and C.~Stricker, \emph{The {Harrison}-{Pliska} arbitrage pricing
  theorem under transaction costs}, J. Math. Econ. \textbf{35} (2001),
  185--196.

\bibitem[LS97]{LevSko97}
S.~Levental and A.V. Skorohod, \emph{On the possibility of hedging options in
  the presence of transaction costs}, Ann. Appl. Probab. \textbf{7} (1997),
  410--443.

\bibitem[MV97]{MerVor97}
F.~Mercurio and T.C.F. Vorst, \emph{Options pricing and hedging in discrete
  time with transaction costs}, Mathematics of Derivative Securities (M.A.H.
  Dempster and S.R. Pliska, eds.), Cambridge University Press, Cambridge, UK,
  1997, pp.~190--215.

\bibitem[Ort01]{Ortu01}
F.~Ortu, \emph{Arbitrage, linear programming and martingales in securities
  markets with bid-ask spreads}, Decis. Econom. Finance \textbf{24} (2001),
  no.~2, 79--105.

\bibitem[PL00]{PerrLef00}
S.~Perrakis and J.~Lefoll, \emph{Option pricing and replication with
  transaction costs and dividends}, J. Econom. Dynam. Control \textbf{24}
  (2000), 1527--1561.

\bibitem[PL04]{PerLef04}
\bysame, \emph{The {A}merican put under transaction costs}, J. Econom. Dynam.
  Control \textbf{28} (2004), 915--935.

\bibitem[Roc97]{Roc97}
R.T. Rockafellar, \emph{Convex analysis}, Princeton University Press,
  Princeton, 1997.

\bibitem[Rou06]{Rou06}
A.~Roux, \emph{European and {A}merican options under proportional transaction
  costs}, Ph.D. thesis, University of York, 2006.

\bibitem[Rut98]{Rut98}
M.~Rutkowski, \emph{Optimality of replication in the {CRR} model with
  transaction costs}, Appl. Math. (Warsaw) \textbf{25} (1998), 29--53.

\bibitem[Sch04]{Sch04}
W.~Schachermayer, \emph{The fundamental theorem of asset pricing under
  proportional transaction costs in finite discrete time}, Math. Fin.
  \textbf{14} (2004), 19--48.

\bibitem[Tok04]{Tok04}
K.~Tokarz, \emph{European and {A}merican option pricing under proportional
  transaction costs}, Ph.D. thesis, University of Hull, 2004.

\bibitem[TZ06]{TokZast06}
K.~Tokarz and T.~Zastawniak, \emph{American contingent claims under small
  proportional transaction costs}, J.~Math. Econom. \textbf{43} (2006), 65--85.

\end{thebibliography}
\end{document}